\documentclass{article}

\usepackage{cite}
\usepackage{listings}
    \usepackage[pdftex]{graphicx}
    \graphicspath{{../figures}}

\usepackage{amsmath}
\usepackage{multirow}
\usepackage{hhline}
\usepackage{lineno}
\usepackage{placeins}
\usepackage[margin=1in]{geometry}
\begin{document}
\title{X Band Fin Resonant Body Transistors in 14nm CMOS Technology}

\author{Jackson~Anderson,
        Yanbo~He,
        Bichoy~Bahr,
        Dana~Weinstein,%
\thanks{Jackson~Anderson (ander906@purdue.edu), Yanbo He, and Dana Weinstein are with the Department
of Electrical and Computer Engineering, Purdue University, West Lafayette, Indiana, USA}
\thanks{Bichoy~Bahr is with Texas Instruments Inc. - Kilby Labs, Dallas, Texas, USA}}

\maketitle

\begin{abstract}

Here we present the first demonstration and in-depth study of unreleased acoustic resonators in 14nm FinFET technology in the IEEE X band, which offer a zero-barrier-to-entry solution for high Q, small footprint, resonant tanks integrated seamlessly in advanced CMOS nodes. These devices leverage phononic waveguides for acoustic confinement, and exploit MOS capacitors and transistors inherent to the technology to electromechanically drive and sense acoustic vibrations. Sixteen device variations are analyzed across thirty bias points to discern the impact of phononic confinement, gate length, and termination scheme on resonator properties. The limiting factor in FinFET resonator performance among design variations tested is shown to be Back End of Line (BEOL) confinement, with devices with acoustic waveguides incorporating Mx and Cx metal layers exhibiting ~2.2x higher average quality factor ($Q$) and peak amplitude, with maximum $Q$ increasing from 115 to 181 and maximum amplitude scaling from 0.8 to 4.5 µS. A detailed analysis of biasing in the highest performing device shows good fit with a derived model, which addresses the velocity saturated piezoresistive effect for the first time in active resonant transistors. Peak differential transconductance that is dominated by changes in the silicon band-structure, as expected from an analysis that includes contributions from the piezoresistive effect, electrostatic modulation, and silicon bandgap modulation.

\end{abstract}

As demand for wireless services grows, so too does the demand for higher frequency radio communication where bandwidth is more plentiful. With 5G FR2 bands extending beyond 10 GHz and FR1 intra-band congestion growing, novel solutions are required to enable high performance, adaptable radios. As a key element of both the filters and voltage controlled oscillators used in radios, high frequency resonators represent an active area of research, with recent developments including thin film bulk acoustic resonators with novel ferroelectric materials, as well as high frequency Lamb and Lamé wave resonators using piezoelectric lithium niobate or aluminum nitride thin films \cite{wang_film_2020,he2020tunable, lu_enabling_2020, assylbekova_11_2020}. To address band congestion, multiple-input-multiple-output (MIMO) arrays are becoming increasingly common as a way to multiplex transmissions between users in the same band, with large arrays allowing for increased performance across end-users sharing spectrum \cite{yang_fifty_2015}. 

The beam-forming enabled by such an antenna array is particularly promising for higher frequency bands, where propagation conditions are less stable due to increased RF absorption. These RF front-ends of increasing complexity require many more analog components to be integrated in a compact form factor, leading to active research in RF integration into traditional digital processes. Many high performance technologies require integration in MEMS-first or MEMS-last process schemes \cite{fedder_technologies_2008, chen_cmos-integrated_2019, riverola_single-resonator_2017}, which provides space and power savings over off-chip RF integration. The use of phononic crystal confinement, however, opens the door for resonator integration directly into a standard CMOS process, potentially enabling larger MIMO arrays in a smaller footprint at lower power \cite{bahr_phononic_2014, bahr_theory_2015}. In addition to reducing parasitics from routing high frequency signals over long distances, this also provides the opportunity to use high performance transistors for active transduction, achieving potentially record-breaking performance \cite{bahr_32ghz_2018}. Such an approach also opens the door for acoustic coupling between devices co-located on the same chip, reducing the need for power hungry phase locked loops and opening the door for exploration of highly scaled coupled oscillator systems for neuromorphic computing \cite{razavi_jitter-power_2021, nikonov_convolutional_nodate}. This paper continues the evolution of such technology by introducing the first CMOS FinFET-based resonators utilizing acoustic waveguiding confinement operating in the IEEE X band (8-12 GHz).

\section{Resonator Design}

\begin{figure*}
	\centering
	\includegraphics[width=0.97\linewidth]{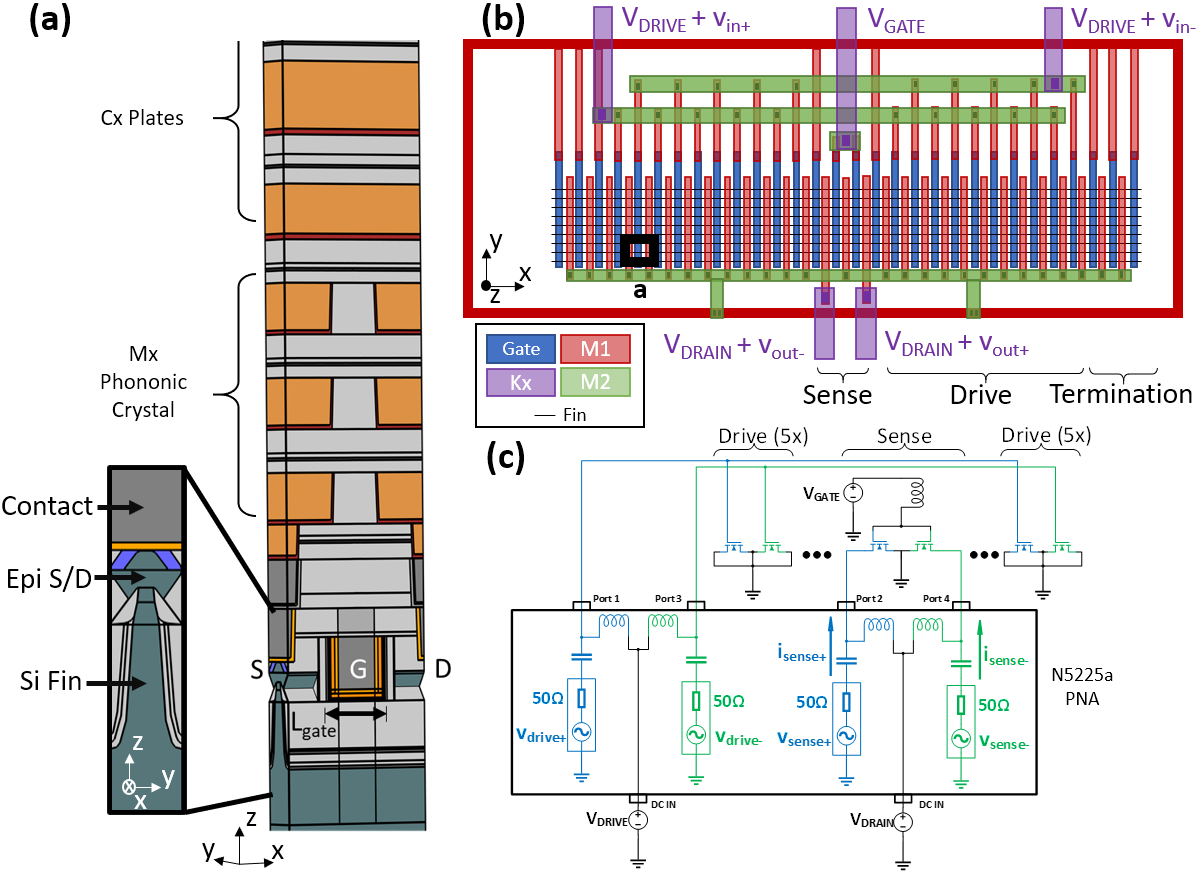}
	\caption{\textbf{(a)} Cross-section of one period of the FinFET resonator , modeled in COMSOL, with inset highlighting Si fin (dark green) and epitaxial raised source and drain with silicide (blue) \cite{noauthor_comsol_nodate}. Metal and contact layers highlighted in orange and dark gray and dielectric layers left uncolored (light gray). The target mode is defined primarily along the x-direction via the 180 degree phase constraint due to differential drive and sense. \textbf{(b)} Top down view of device shows the central sense transistors with five pairs of capacitors on either side allowing for differential drive, capped on either end by termination (ten gates in the actual devices, abbreviated to three in the figure) and surrounded by a ground ring. First level metal M1 is used for routing within the structure while higher level metals in Mx and Cx (not shown for clarity in this diagram) rest above this routing in either continuous plate or PnC form. The Kx layer represents higher level metal layers used for device to pad routing. \textbf{(c)} Resonator electrical test scheme with DC biasing provided by standalone SMUs, routed through internal PNA bias tees.  }
	\label{fig:deviceoverview}
\end{figure*}

The designed resonator structure, shown in Fig. \ref{fig:deviceoverview}, is fully integrated in the GlobalFoundries 14LPP process and occupies a space of 9 µm x 15 µm \cite{globalfoundries_12lp_2018}. The device is a four port network, with output of the resonator taken as the differential transconductance (Eq. \ref{eq:gmdd}), which is a function of three main items: the electrical to acoustic transduction efficiency in the drive transistors, the acoustic energy confinement and amplification in the resonance cavity, and the acoustic to drain current conversion in the output transistors. The first of these items, drive transduction, is electrostatic due to the 14 nm fabrication process used, although the potential for future piezoelectric drive using CMOS compatible thin film ferroelectrics exists \cite{trentzsch_28nm_2016,he2020tunable,he2019switchable}. The third of these, acoustic-to-drain current conversion, is also largely a function of the process and materials used in device fabrication. While this conversion will be examined utilizing a simplified model, the initial device design is focused on maximizing acoustic energy confinement by varying gate length, termination scheme, and back-end-of-line (BEOL) confinement conditions.

\begin{equation}\label{eq:gmdd}
	g_{mdd} = \frac{i_d}{v_d} = Y_{dd21}-Y_{dd12}
\end{equation}

\subsection{BEOL Phononic Confinement}

Fig. \ref{fig:deviceoverview}(c) depicts a 3D unit cell of the resonator with a 14 nm wide FinFET at the front-end-of-line (FEOL) and periodically arranged metal blocks at the BEOL, categorized into Mx and Cx layers as shown based on thickness and minimum lateral feature sizes. The target resonant mode for the unreleased resonator is along the length direction of the fin transistors. To obtain lower insertion loss and higher quality factor, the radiated acoustic energy needs to be concentrated at the MOS capacitor drive transducers and the sensing transistors. Compared to more traditional suspended resonators, this represents a challenge in the unreleased resonant body transistor design in the commercial CMOS processes.

Integrating a phononic crystal (PnC) at the BEOL of the standard CMOS process and leveraging the mechanical bandgaps of such structures has proven to be an effective tool to confine the energy and boost resonance quality factor \cite{he2020tunable,bahr_theory_2015}. As shown by M1 in \ref{fig:deviceoverview}(b), these periodic metals are uniform along the length (y direction) of the transistor gates (orthogonal to the fin direction) within the resonant cavity. 

In order to study and optimize the bandgap induced by the phononic crystals, several 2D simulations were performed in COMSOL Multiphysics. Specifically, by mapping the real coordinates to reciprocal coordinates and searching for the eigenmodes along the first irreducible Brillouin zone in the 2D reciprocal lattice, a 130 nm by 60 nm PnC (such as can be designed within the constraints of the Mx layers in the 14 nm process) yields a calculated bandgap from 8 GHz to 12 GHz (shown in Fig. \ref{fig:DR}(b)). 
To further confine the elastic energy, a separate PnC designed under Cx metal layer design rules is also considered for a subset of devices. The corresponding dispersion relationship is shown in Fig. \ref{fig:DR}(c). A mechanical bandgap is then obtained between 7.6 GHz and 10.7 GHz which overlaps with the bandgap induced by the Mx metal PnC.
Together, these PnC designs indicate elastic waves with a frequency of 8 GHz to 12 GHz can be prohibited from propagating into the BEOL. If coupled with a waveguided mode with differential drive that exhibits a large momentum ($k_x = \pi/a$) at a frequency below the sound cone for the bulk silicon substrate, these modes can be confined laterally towards the driving/sensing transistors, with modes from 8 to 10.7 GHz showing the greatest confinement \cite{bahr_theory_2015}. For an in-depth analysis on the acoustic modes in this device, readers are referred to \cite{rawat_analysis_2021}.

\subsection{Resonant Cavity Termination}
 While the PnCs in the main resonant cavity enhance energy confinement and wave reflection from the BEOL, there also exists wave scattering at the two ends of the resonator. Reducing this scattering further increases the horizontal confinement, reduces the insertion loss, and hence increases the Q. The adiabatic theorem in photonic waveguide design is leveraged here, which states that when the wave is propagated down the waveguide, scattering will vanish because of the limitation of sufficiently slow perturbation \cite{meade1995photonic}. Therefore, the key to reduce this scattering is to introduce different units as terminations at the two ends for slower transitions. The unit cell for all the PnC structures in the major termination section throughout the paper is kept as 150 nm x 60 nm.
 
 While all measured devices have ten terminating gates at each end of the cavity, two types of cavity termination spacings are explored to provide acoustic confinement in the lateral direction. These include abrupt and gradual termination schemes. The abrupt termination refers to the scenario where a terminating periodic array with constant pitch shifted from that of the resonance cavity array is immediately adjacent to the main resonance region. In this case, the gate length of the termination immediately transitions to the $L_{term}$ value given in Fig. \ref{fig:rfoverview}(a) and stays  constant until the end of the device. As such, the guided waves are abruptly transitioned from the main cavity towards the ends. In the gradual termination scenario, the main resonant cavity periodicity adiabatically transitions from $L_{gate}$ to the terminating array periodicity $L_{term}$, resulting in a termination gate length gradually transitioning from 80 to 140 nm. This serves to reduce scattering from the two ends and boost the Q \cite{bahr2015theory}. These two schemes are depicted in Fig. \ref{fig:rfoverview}(c).

\subsection{FinFET Sensing}

To understand the conversion from stress to drain current in the sense transistors, modulation of three separate FET properties is examined: oxide capacitance, channel mobility due to piezoresistivity, and silicon bandgap. 
For ease of visualization, an initial analysis is performed with the source-referenced simplified strong-inversion model with $\alpha=1$ (the traditional "square-law model", or SPICE level 1) \footnote{This model has only a handful of parameters to describe device operation, as opposed to the more advanced BSIM models used commonly in industry, which include many empirically fit parameters to give the best device models for commercial applications \cite{noauthor_bsim_nodate}. This simplified model is used with the understanding that the derivation, while not capturing a variety of short channel effects that influence modern transistor behaviour, will provide a more intuitive understanding into the basics of active FET sensing.} \cite{tsividis_operation_2010}:

\begin{equation}
	I_d = \mu C_{ox} \frac{W}{L}\Big((V_{GS}-V_T)V_{DS} - \frac{V_{DS}^2}{2}\Big)
\end{equation}
where $I_d$ is the drain current, $\mu$ is the carrier mobility, $C_{ox}$ is the gate capacitance, $V_{DS}$ and $V_{GS}$ are the drain-to-source and gate-to-source bias, respectively, $W$ and $L$ are the channel width and length, and $V_T$ is the threshold voltage required for significant channel conduction.

Capacitance modulation in these devices is the same mechanism used in electrostatic drive of these devices, except that rather than sensing the gate current directly, the capacitance modulation is being amplified through transistor action into a change in drain current. The piezoresistive effect, whereby the semiconductor carrier mobility is modulated through stress in the transistor channel, likewise appears as a multiplier on the transistor drain current. Semiconductor bandgap modulation, on the other hand, largely effects the threshold voltage of the transistor. For a complete derivation of these effects, the reader is directed to the derivation in Appendix \ref{sec:derivation}.
Modified to include the effects of stress on $\mu$, $C_{ox}'$, and $V_T$ (modeled as $\mu(\sigma) = \mu_0+\Delta \mu$), the original expression for current in the linear region becomes the following:

\begin{multline} \label{eq:id_stressfet}
	I_d(\sigma)	= (\mu_0 C_{ox0}' +\mu_0 \Delta C_{ox}' + C_{ox0}'\Delta \mu + \Delta \mu\Delta C_{ox}')  *\frac{W}{L}\Big((V_{GS}-V_{T0})V_{DS} - \frac{V_{DS}^2}{2}\Big) \\
	-\Delta V_T V_{DS} \frac{W}{L} (\mu_0 C_{ox0}' +\mu_0 \Delta C_{ox}' + C_{ox0}'\Delta \mu + \Delta \mu\Delta C_{ox}')
\end{multline}
where
\begin{multline} \label{eq:deltaid_stressfet}
	\Delta I_d(\sigma)	= (\mu_0 \Delta C_{ox}' + C_{ox0}'\Delta \mu + \Delta \mu\Delta C_{ox}') *\frac{W}{L}\Big((V_{GS}-V_{T0})V_{DS} - \frac{V_{DS}^2}{2}\Big) \\
	-\Delta V_T V_{DS} \frac{W}{L} (\mu_0 C_{ox0}' +\mu_0 \Delta C_{ox}' + C_{ox0}'\Delta \mu + \Delta \mu\Delta C_{ox}')
\end{multline}

Likewise, in saturation

\begin{equation}
	I_{d0} = \mu C_{ox} \frac{W}{2L}\Big((V_{GS}-V_T)^2\big(1+\lambda[V_{DS}-(V_{GS}-V_{T0})]\big)\Big)
\end{equation}
where $\lambda$ represents the channel length modulation coefficient (0.05 to 0.25 typical) and 

\begin{multline}
	\Delta I_d 
	= (\mu_0 \Delta C_{ox}' + C_{ox0}'\Delta \mu + \Delta \mu\Delta C_{ox}')
	*\frac{W}{2L}\Big((V_{GS}-V_T)^2(1+\lambda(V_{DS} - V_{GS} + V_{T}))\Big) \\
	\hspace{13mm}- \frac{W}{2L} (\mu_0 C_{ox0} +\mu_0 \Delta C_{ox} + C_{ox0}\Delta \mu + \Delta \mu\Delta C_{ox}) \Bigg[ \hfill \\ 
	\hspace{27mm}\Delta V_T \Big[\lambda(V_{GS}-V_{T0})^2 - 2(V_{GS}-V_{T0}) 
	*\Big(1+\lambda\big[V_{DS}-(V_{GS}-V_{T0})\big]\Big)\Big] \hfill \\ 
	\hspace{23mm}+ \Delta V_T^2 \Big[\Big(1+\lambda\big[V_{DS}-(V_{GS}-V_{T0})\big]\Big) - 2\lambda(V_{GS}-V_{T0})\Big] \hfill \\
	\hspace{23mm}+ \Delta V_T^3 [\lambda]  \Bigg] \hfill 
\end{multline}

For modern sub-micron devices, carriers do not exhibit constant mobility but rather undergo velocity saturation at high fields. A simplified way to model this analytically is through a piece-wise equation
\begin{equation}
	v_d = \begin{cases}
		\frac{\mu |E_x|}{1+|E_x|/E_{sat}} \quad,\quad |E_x| < E_{sat} \\
		v_{sat}\hfill,\quad |E_x| > E_{sat}
	\end{cases}
\end{equation}
which, solved at $v_d = v_{sat}$, gives
\begin{equation} \label{eq:esat}
	E_{sat} = \frac{2v_{sat}}{\mu}
\end{equation}
and leads to a modified expression for drain current
\begin{equation}
	I_d = \begin{cases}
		\mu C_{ox} \cfrac{W}{L}\cfrac{1}{1+\cfrac{V_{DS}}{E_{sat}L}}\Big((V_{GS}-V_T)V_{DS} - \frac{V_{DS}^2}{2}\Big) \quad,\quad V_{DS} < V_{DSsat} \\ \\
		C_{ox} W v_{sat}\cfrac{(V_{GS}-V_T)^2}{(V_{GS}-V_T)+E_{sat}(L-\Delta L)} \hfill,\quad V_{DS} > V_{DSsat}
	\end{cases}
\end{equation}
where
\begin{equation}
	V_{DSsat} = \frac{(V_{GS}-V_T)E_{sat}L}{(V_{GS}-V_T)+E_{sat}L}.
\end{equation}

An important nuance for acoustic resonators is that both the saturation velocity and the change in mobility (piezoresistive effect) are related to the change in carrier effective mass caused by acoustic deformation of the crystal lattice - saturation velocity by $m^{*-1/2}$ and mobility by a factor between $m^{*-1/2}$ and $m^{*-5/2}$, depending on doping (for a detailed analysis, see Appendix \ref{sec:mobility}). This relation means that, while DC current may saturate, the AC piezoresistive effect can still change drain current to an extent, even in the saturation region. For this reason, a compound model is utilized that incorporates velocity saturation in the calculation of DC drain current but allows mobility to vary freely. Under this approach, the contribution to RF output from both electrostatic forces and mobility modulation should scale linearly with drain current. The contribution from any threshold voltage shift is seen to be related to drain bias only in the linear regime and is scaled by the impact of mobility and capacitance changes. In the saturation regime, on the other hand, the threshold voltage shift is primarily a function of gate bias and is modified by drain bias only through channel length modulation. These expressions for $\Delta I_d$ represent one half of the device transfer function $g_{mdd}$, discussed in Eq. \ref{eq:gmdd}, with the other half being given by the voltage-to-strain conversion and subsequent strain confinement/resonance. While direct measurement of strain in the solid-state resonator in the front end of line poses a challenge, all three of these aspects are subsequently investigated using measured electrical results.

\section{Results and Discussion}

DC results, shown as an example for the 80 nm gate length device A1 in Appendix \ref{sec:sup_figs}, indicate that the sense transistors in all devices were working as intended. The large current values of approximately 2 mA when gate and drain are biased at $V_{dd}$ are due to the parallel DC connection of the two sense transistors and large effective width of the individual transistors, where 40 fins span the resonance cavity (in y-direction) and are electrically connected in parallel. 

\subsection{Analysis of Design Variations}

\begin{figure*}
	\centering
	\includegraphics[width=1\linewidth]{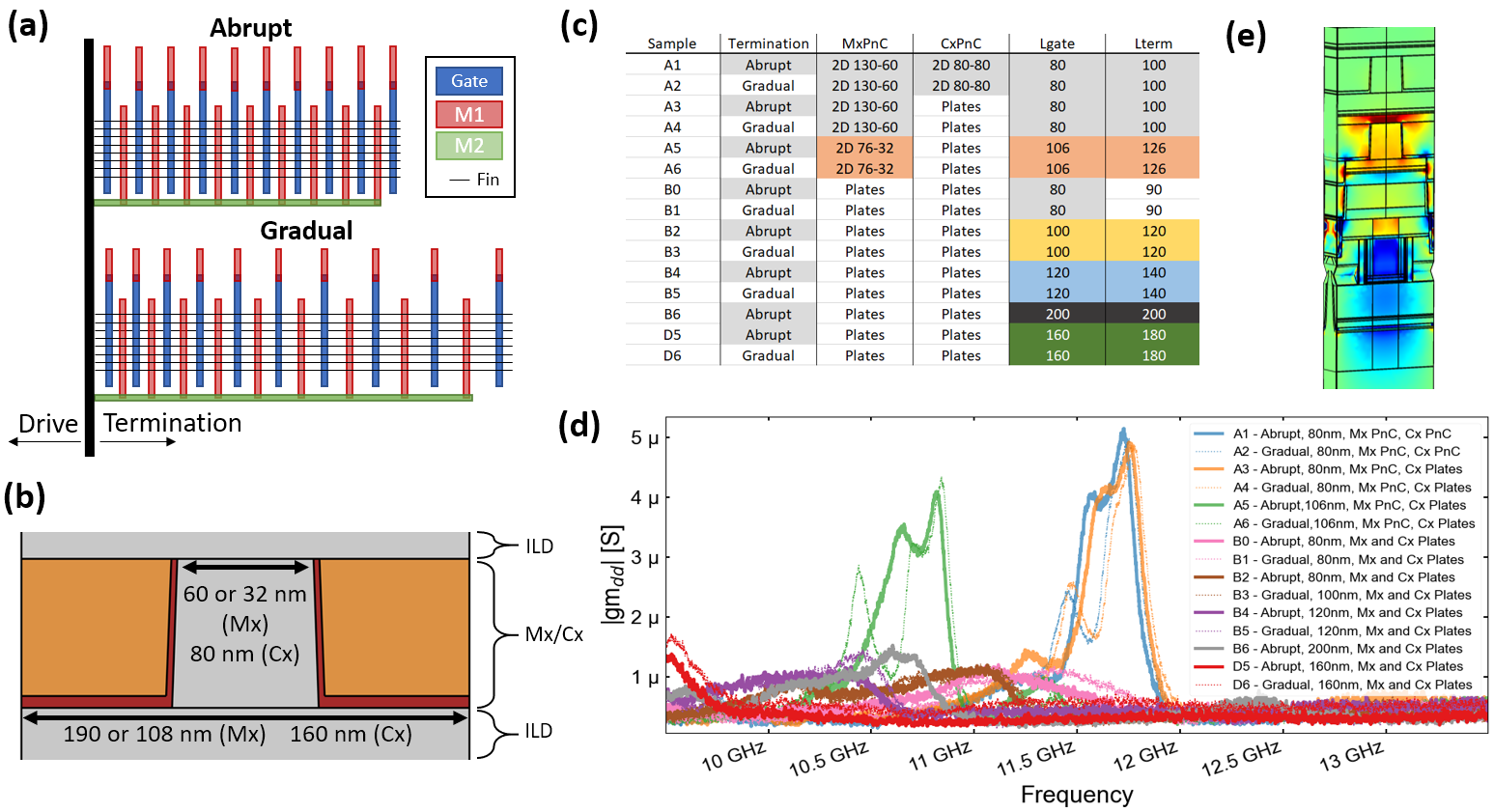}
	\caption{\textbf{(a)} Depiction of the termination schemes investigated for use at the ends of the resonant cavity. \textbf{(b)} Phononic crystal structure, with various investigated pitches. \textbf{(c)} Device design variations and \textbf{(d)} their experimentally measured differential transconductance at $V_D=0.6$ V, $V_G=0.8$ V, $V_{DR}=0.8$ V. \textbf{(e)} Target mode, simulated in COMSOL, with stress concentrated in finFET channel and confined by the BEOL metal layers.}
	\label{fig:rfoverview}
\end{figure*}

While DC results were similar for all devices (outside expected differences due to gate length), resonance mode confinement was highly dependent on the design variations studied. A summary of all measured devices as well as a comparison of their differential transconductance spectra from 9.5 to 13.5 GHz are shown in Fig. \ref{fig:rfoverview}. From these data, it is evident that phononic crystal confinement in the Mx layers results in significantly improved quality factor and maximum transconductance. Also visible is the role the type of termination has in defining the center frequency of nearby spurious modes in the device. 

\begin{table*}[]
	\renewcommand{\arraystretch}{1.3}
	\caption{ANOVA Results for Peak Fits}
	\label{tab:anova_results}
	\centering
	\resizebox{\textwidth}{!}{%
	\begin{tabular}{|c|c|c||c|c|c|c|c|c||c|}
		\hline
		\multicolumn{2}{|c|}{\scriptsize{95\% Confidence Intervals} } & \multirow{2}{*}{n} & \multicolumn{2}{c|}{Frequency [Hz]} & \multicolumn{2}{c|}{Height [nS]} & \multicolumn{2}{c||}{Q} & \multirow{2}{*}{Samples} \\
		\cline{4-9}
		\multicolumn{2}{|c|}{\scriptsize{Significant results in bold.}} &  & Mean & Std. Error & Mean & Std. Error & Mean & Std. Error &  \\
		\hhline{|==|=||=|=|=|=|=|=||=|}
		\rule[-1ex]{0pt}{2.5ex} \multirow{3}{*}{PnC Confinement} & Mx, Cx & 413 & 11.27 G & 52.8 M & \textbf{451.1} & \textbf{30.6} & \textbf{44.32} & \textbf{1.68} & \multirow{3}{*}{\shortstack{A1-A4, B0-B1\\ (L$_{gate}$=80 nm)}} \\
		\cline{2-9}
		\rule[-1ex]{0pt}{2.5ex}  & Mx Only & 368 & 11.22 G & 55.9 M & \textbf{379.4} & \textbf{32.5} & \textbf{37.55} & \textbf{1.78} &  \\
		\cline{2-9}
		\rule[-1ex]{0pt}{2.5ex}  & Plates & 276 & 11.26 G & 64.6 M & \textbf{203.6} & \textbf{37.5} & \textbf{19.88} & \textbf{2.05} &  \\
		\hline
		\rule[-1ex]{0pt}{2.5ex} \multirow{2}{*}{Termination} & Abrupt & 1289 & \textbf{11.16 G} & \textbf{33.4 M} & \textbf{301.6} & \textbf{13.9} & \textbf{24.21} & \textbf{0.795} & \multirow{2}{*}{All} \\
		\cline{2-9}
		\rule[-1ex]{0pt}{2.5ex}  & Gradual & 1145 & \textbf{11.29 G} & \textbf{35.4 M} & \textbf{352.4} & \textbf{14.7} & \textbf{ 31.71} & \textbf{0.843} &  \\
		\hline
		\rule[-1ex]{0pt}{2.5ex} \multirow{2}{*}{L$_{gate}$} & 80 nm & 30 & \textbf{11.66 G} & \textbf{23.0 M} & 1,866 & 153 & 84.8 & 4.40 & \multirow{2}{*}{\shortstack{A3-A6 \\ Q$>$25, Height$>$500 nS}} \\
		\cline{2-9}
		\rule[-1ex]{0pt}{2.5ex}  & 106 nm & 30 & \textbf{10.71 G} & \textbf{23.0 M} & 1,651 & 153 & 87.1 & 4.40 &  \\
		\hline
	\end{tabular}}
\end{table*}

These differences are quantified in Table \ref{tab:anova_results}, where the center frequency, quality factor, and height of resonant peaks in $g_{mdd}$ are compared for the three design splits after elimination of high-error peak fit ($\approx$5\% of data overall). For termination, this process is straightforward as there exists an abrupt and gradually terminated copy of every device. The resulting analysis indicates an increased transconductance in gradual transition devices (352 nS) vs. abruptly terminated ones (302 nS), with a corresponding increase in average quality factor (32 vs. 24). In addition to this enhancement, average resonance center frequency was shown to decrease slightly from 11.29 to 11.16 GHz, driven by the downward shift in the left-most peak of the high-$Q$ cluster of modes seen in PnC-confined devices. 

\begin{figure*}
	\centering
	\includegraphics[width=0.9\linewidth]{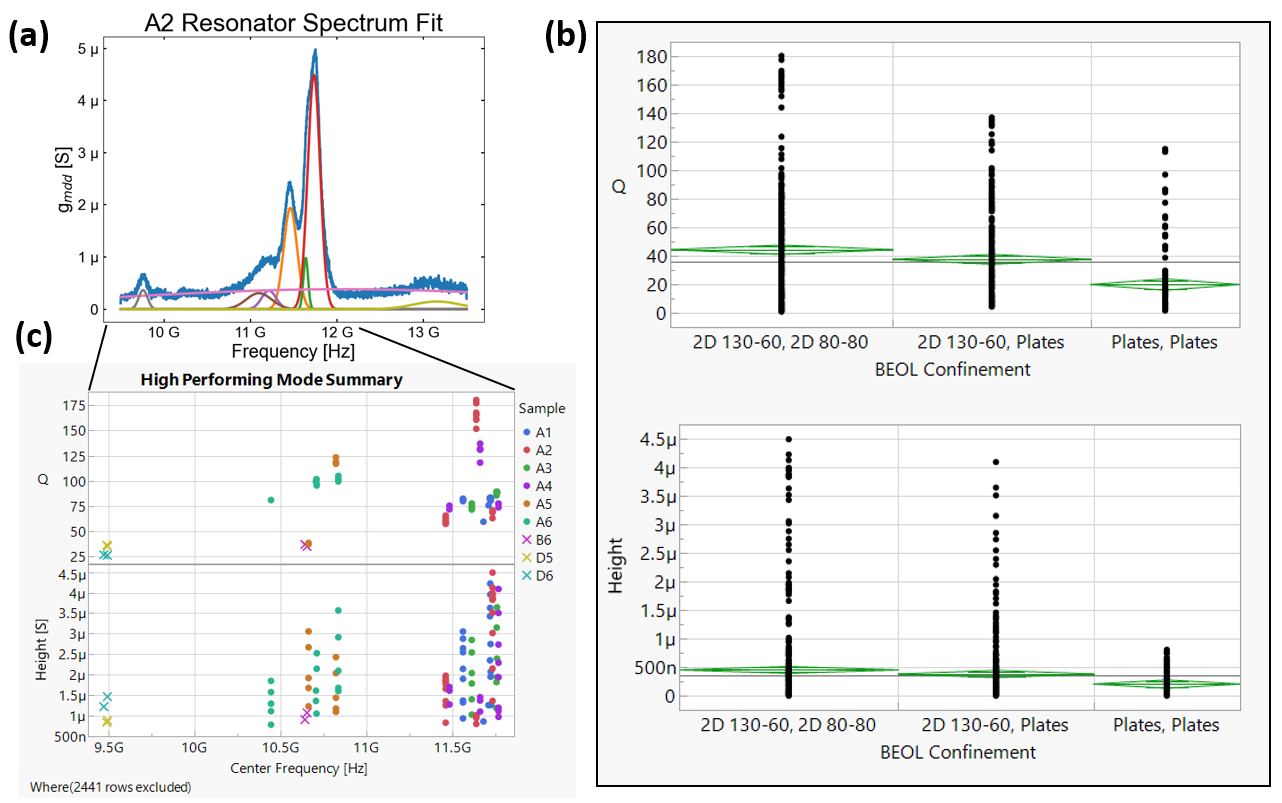}
	\caption{ \textbf{(a)}: Output of peak fit algorithm for one of the 392 individual measurements evaluated. Peak fits with error above 0.025 were filtered (120 out of 2554 peaks) and then design splits were evaluated by comparing means and performing ANOVA between appropriate samples. The most significant process variation, BEOL confinement, is shown in \textbf{(b)}, with complete results given in Table \ref{tab:anova_results}. In addition to increasing average peak height and $Q$ across a range of drive and drain biases, it is also evident that an Mx-PnC greatly increases the maximum peak height, from under 1 µS to 4.5 µS for the highest-performing resonator. \textbf{(c)} shows Q and amplitude of fit resonance peaks, filtered by a minimum $Q$ of 25 and height of 750 nS, with each dot indicating a peak of the correspondingly colored device under a single bias condition. Transconductance variations at the same frequency show the dependence of the resonance on electrical biases. Also evident is the change in center frequency due to gate length, and splitting from two to three distinct modes with gradual termination, as first shown in Fig. \ref{fig:rfoverview}. }
	\label{fig:peakfitresults}
\end{figure*}

To investigate phononic crystal confinement, analysis was limited to designs with 80 nm gate length (A1-A4 and B0-B1) and the confinement treated as a three-level categorical factor. This analysis showed no significant change in center frequency of the modes, but it did show significantly improved transconductance (204, 379, 451 nS) and quality factor (20, 38, 44) with increasing layers of phononic crystal confinement. These trends are highlighted in Fig. \ref{fig:peakfitresults} along with an example of the peak fitting results used for parameter extraction.

Compared to the other factors studied, L$_{gate}$ is unique in that it is the primary design variable for selecting center frequency due to the mode being defined by source/drain spacing. This means that the different values studied dramatically change the cavity shape and thus all peaks detected in the studied frequency span of 9.5 to 13.5 GHz cannot be used in the analysis, as they are not present in this range in all devices. To overcome this, a much smaller comparison of A3 and A4 (L$_{gate}$=80 nm) to A5 and A6 (L$_{gate}$=106 nm) was performed using only the prominent modes with transconductance above 500 nS and $Q$ above 25 that were present towards the center of the spectra. The downside of this approach is that gate length is confounded with a change in phononic crystal design that is necessary for the different device geometries due to design rule challenges. Despite this, the fact that confinement method showed no significant change in center frequency for 80 nm devices points towards this shift being dominated by gate length as would be expected.

\begin{figure*}
	\centering
	\includegraphics[width=1\linewidth]{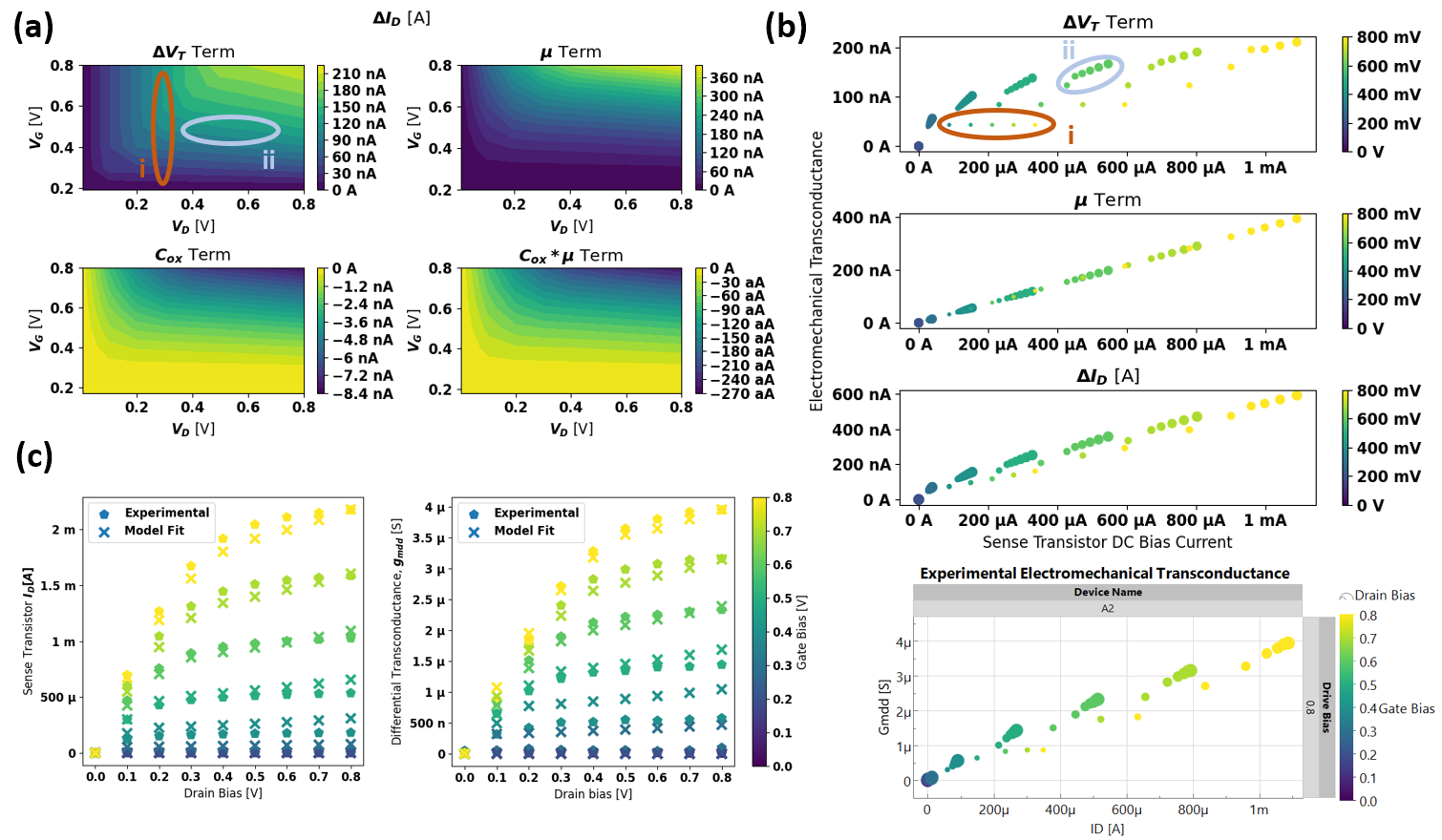}
	\caption{\textbf{(a)} Simulated contributions to $\Delta I_D$ from bandgap modulation (threshold voltage), piezoresistive modulation (mobility), electrostatic modulation (oxide capacitance), and cross-term. For short channel devices that are supply voltage constrained, bandgap and piezoresistive modulation are of similar magnitudes and dominate over electrostatic terms.  \textbf{(b)} These two terms, along with the overall drain current modulation, are plotted against DC bias current of the sense transistor (X-axis) for simulated (top) and experimental (bottom) devices. Gate bias is shown by marker color (blue to yellow) and ranges from 0.4 to 0.8 by 0.1 V. Drain bias is represented with marker size, moving from small to large as bias increases from 0.1 to 0.8 by 0.1 V. The large cluster of points with equal gate biases and similar drain currents indicate saturation mode operation for those bias conditions with channel length modulation. The mobility term is linear with drain current, whereas the threshold voltage term due to bandgap modulation has both a flat \textbf{(i)} region in the linear regime with no $V_G$ dependence for constant $V_D$ and a slight $V_D$ dependence \textbf{(ii)} for constant $V_G$ in the saturation regime. Experimental data shows roughly linear dependence of drain current on gate voltage (a hallmark of velocity saturation \cite{tsividis_operation_2010} vs. the $V_G^2$ dependence of the long channel model). Some deviation between the model fit and experimental results can be seen at high drain bias and low gate bias, which may be indicative of mobility degradation due to vertical fields, which is not modeled currently. Otherwise, experimental results show a similar shape to that of the simulated device, indicating the dominance of both piezoresistive and bandgap modulation in the experimental resonator. \textbf{(c)} Model fit to experimental results ($\mu=79$ cm$^2$/Vs, equivalent oxide thickness=1.6 nm, $V_t$=0.206 V, $v_{sat}$=1.83*10$^7$ cm/s, $\lambda$=0.47, $W$ = 3 um, total stress=1 MPa). A basin-hopping algorithm is used to find a global minimum fit for the DC transistor parameters, which are then fixed for the AC fit, achieved by varying stress.}
	\label{fig:exp_vs_sim}
\end{figure*}

As the device with the highest performance, resonator A2 was selected for further analysis, with the maximum transconductance of the $\approx$11.75 GHz mode examined as a function of device biasing. This was done at a drive bias of $V_{dd}$, as $g_{mdd}$ was seen to increase linearly with the increase in MOS capacitance at higher drive bias (see Appendix \ref{sec:sup_figs}). The dependence of $g_{mdd}$ on sense transistor biasing, shown in Fig. \ref{fig:exp_vs_sim}, highlights that device transconductance, as expected, is modulated in a manner consistent with piezoresistive and bandgap modulation. While transistor parameters are of the correct order of magnitude in the reported data fit, they vary slightly from real-world values due to the simplified model used. For example, the square law model does not accurately capture subthreshold (accumulation and weak inversion) behaviour, and thus the fit favors a slightly lower Vt to more closely fit bias currents near zero. This, in turn, leads to an overestimation of current at higher biases, which is compensated by a lower than expected value for mobility and a slightly larger equivalent oxide thickness. On the AC side, the fit is hindered by the presence of a noise floor in the s-parameter measurements that may obscure lower peak levels. This, along with the poor near-threshold characteristics of the square law model, contributes to the fitting error at lower values of $g_{mdd}$. Moving away from the square law model to a more advanced model such as that derived by Taur et al. that captures all regions of operation may decrease this error \cite{taur_continuous_2004}. Likewise, adopting band-structure simulations of silicon as a result of acoustic deformation could provide a unified source of mobility and threshold voltage changes in the transistor, at the expense of increased computation requirements.

For NMOS devices oriented in the <110> direction, the negative piezoresistive coefficients ensure that the contribution of the threshold voltage shift and mobility is additive. If fRBT devices are implemented with PMOS devices or in processes with different fin orientations, care should be taken to ensure that the dominant stress-dependent effects do not compete against each other. This may not be as much of an issue for long channel RBT devices, as piezoresistive modulation dominates for long channel devices that are not supply voltage constrained. 

The highest magnitude peak fit occurred in device A2 (L$_{gate}$ of 80 nm, gradually terminated, dual PnC) at 11.73 GHz with a differential transconductance of 4.49 µS and a $Q$ of 69.8, for an $f*Q$ product of 8.19*10$^{11}$ at drain, gate, and drive biases of 0.6, 0.8, and 0.8 V, respectively. While previous unreleased resonators in the 32 SOI process have achieved an $f*Q$ product of $3.8*10^{13}$ with phononic confinement at 3.3 GHz, these 14 nm devices exhibit 50x higher transconductance and show an almost 3x improvement in Q over a similar frequency measured mode in 32 SOI devices \cite{bahr_optimization_2016, marathe_resonant_2014}. Compared to some contemporary resonators in the X and adjacent K$_\textrm{u}$ (12-18 GHz) bands in Table \ref{tab:lit_comparison}, the unreleased devices in this work achieve moderate $f*Q$ in a commercially available process in an area smaller than comparable released devices. This, combined with the fact that the devices are integrated in the front-end-of-line of a commercial CMOS advanced technology node, opens the door for tight logic integration with minimum routing (and associated parasitics). Additionally, if process design rule challenges can be overcome, this performance can be achieved across a band of frequencies by varying gate length and phononic crystal design.

\begin{table*}
	\renewcommand{\arraystretch}{1.3}
	\centering
	\caption{Comparison with Contemporaneously Reported X and K$_\textrm{u}$ Band Resonators}
	\label{tab:lit_comparison}
	\begin{tabular}{|c|c|c|c|c|c|c|}
		\hline
		Ref & $f_0$ [GHz] & $Q$ & $f_0 * Q$ & Technology & Released & Area [um$^2$] \\
		\hline
		\textbf{This Work} & 11.7 & 69.8 & 8.2*10$^{11}$ & 14LPP & No & 135 \\
		\hline
		\cite{marathe_resonant_2014} & 11.5 & 24 & 2.8*10$^{11}$ & 32SOI  & No & 15 \\
		\hline
		\cite{chen_high-q_2019} & 8.8 & 750 & 6.6*10$^{12}$ & Custom AlN & Yes & 6,500 \\
		\hline
		\cite{assylbekova_11_2020} & 11.1 & 615 & 6.8*10$^{12}$ & Custom AlN & Yes & 675 \\
		\hline
		\cite{yang_1060-ghz_2020} & 13.0 & 282 & 3.7*10$^{12}$ & Custom LiNbO3 & Yes & 2,000 \\
		\hline
		\cite{abouyoussef_quad_2018} & 14.5 & 400 & 5.8*10$^{12}$ & PCB Microstrip  & - & 27,000,000 \\
		\hline
	\end{tabular}
\end{table*}

\section{Conclusion}

This work provides an in-depth experimental investigation into the influence of several design parameters on the performance of unreleased Resonant Fin Transistors in standard 14nm FinFET technology. Phononic crystal confinement, shown experimentally here to be the most important variation in improving unreleased resonator performance (2.2x improvement on average), provides significant opportunity for integrating acoustic devices into standard CMOS platforms. The highest performing resonator, a gradually terminated 80 nm gate length device with both Mx and Cx phononic crystal confinement, shows a transconductance amplitude and $Q$ of 4.49 µS and 69.8 respectively, with an $f*Q$ product of 8.19*10$^{11}$. At the same time, an analytical model has been demonstrated that predicts resonator performance across sense transistor biasing conditions, utilizing a combined velocity saturated current at DC and unsaturated mobility variation at AC, where acoustic perturbation of carrier effective mass modulates saturation velocity and mobility simultaneously. This highlights the importance of understanding the assumptions that go into developing advanced electrical models for short channel transistors, which may not accurately depict effects present in active electromechanical devices. While creating these phononic confinement schemes in a standard process with strict design rules poses a unique challenge, we have established a methodical way to predict and design resonator performance across multiple CMOS platforms. This technology combined with parallel advancements in CMOS-compatible ferroelectric and piezoelectric thin films, presents the opportunity to develop compact, low-cost, CMOS-integrated, and electrically-controllable resonators with no additional packaging required, opening the door for acoustic processing alongside traditional electronics in advanced CMOS nodes.

\appendix
\section{Characterization Methods}

\subsection{Device Measurement}

S-parameter measurements were taken at room temperature and pressure, with an input power of -10 dBm using an Agilent Technologies N5225a PNA, with biasing provided by three Keithley
2400 SMUs as shown in Fig. \ref{fig:deviceoverview}a. All four devices were connected via GPIB and synchronized via published Python scripts  \cite{anderson_pymeasrf_2019, anderson_pymeasrf_2020}.
Wincal software from Formfactor Inc. was used to facilitate pre-measurement calibration, with a Hybrid LRRM-SOLR
performed utilizing an Impedance Standard Substrate (ISS 129-246).
Infinity GSSG Probes were used to allow for a compact biasing scheme, saving die real estate at the expense of
increased cross-talk between adjacent signal lines.
Gate bias for the sense transistors was provided via a separate DC needle probe.

\subsection{Measurement Data Post-Processing}

Acquired Touchstone files were first analyzed using Scikit-RF, which was used to transform the single-ended
parameters collected into their mixed mode equivalent matrices \cite{scikit-rf_developers_scikit-rf_2020}.
From these, gmdd was extracted to allow for analysis of the designed differential mode.
The magnitude of $g_{mdd}$ for each sample at a bias of V$_{drain}=0.6$, V$_{gate}=0.8 $, and V$_{drive}=0.8$ V, was modeled as a collection of Gaussian peaks and fit via
lmfit using the Levenberg-Marquardt algorithm with initial guesses for frequency and peak height provided via user input
through matplotlib's ginput method \cite{noauthor_lmfit_2020}.
These preliminary fits were then used as initial conditions for the remainder of the 377 collected measurements taken at drain biases of 0.2, 0.4, and 0.6 V, drive biases of 0, 0.2, 0.4, 0.6, and 0.8 V, and sense FET gate biases of 0 and 0.8 V, These were
again fit via lmfit and parallelized using dask \cite{noauthor_dask_2020}. Due to the noisy nature of the data, with varying spectra between devices, a number of constraints were added to improve fitting accuracy over all design variations. The fitting was weighted by gmdd amplitude to emphasize primary peaks over smaller spurious modes and background noise. Variation in the peak standard deviation between biases was constrained to two times the original standard deviation. Center frequency for each sample was constrained to a maximum of 5\% deviation as a function of bias to improve fitting accuracy and reflect the fact that mode shape, and thus frequency of operation, is dictated primarily by device geometry.

Extracted peak properties were then analyzed in SAS JMP statistical discovery software to investigate the relationship
between device design splits and variation in performance as a function of device bias \cite{noauthor_jmp_2019}. The device splits, shown in Fig. \ref{fig:rfoverview}, are not entirely orthogonal due to a variety of device design constraints related to the phononic crystals and gate lengths. Thus, one-way analysis of variance (ANOVA) was used on subsets of the experimental devices rather than trying to fit a multidimensional model to the entire sample set. 

Examination of expected device performance using the derived analytical model was performed in Python. Contribution for each of the three components (bandgap, mobility, and capacitance modulation) was taken as the difference between the positive and negative stress components. Lmfit was used to perform the optimization routine, using a basinhopping algorithm to find a global minimum for the DC current. The fit of $g_{mdd}$ was performed using the Levenberg-Marquardt algorithm as this had less tendency to settle to a poor-fitting local minima. The simplified 1D model of a 3D transistor led to significant cross-correlation between stress directional components, however magnitude of the total stress remained relatively constant across fit iterations.

\section{FinFET Sensing: Electromechanical Conversion from Stress to Drain Current} \label{sec:derivation}

To understand the conversion from stress to drain current in the sense transistors, modulation of three separate FET properties is examined: channel mobility due to piezoresistivity, oxide capacitance, and silicon bandgap.

\subsection{Mobility}
The piezoresistive effect is typically modeled as a linear change in resistivity ($\rho$) as a function of applied stress ($\sigma$) and the longitudinal and transverse piezoresistive coefficients ($\pi$):
\begin{equation}
	\rho = \rho_0[1+\pi_L\sigma_L+\pi_T\sigma_T].
\end{equation}
Low-field mobility ($\mu$) is related to the resistivity as follows, where q is the elementary charge and N is the doping concentration per unit volume.
\begin{equation}
	\mu = 1/q\rho N
\end{equation}
A stress-dependent mobility can be defined in terms of the initial mobility $\mu_0$,
\begin{equation}
	\mu(\sigma) = \frac{\mu_0}{1+\pi_L\sigma_L+\pi_T\sigma_T}
\end{equation}
and the initial value of mobility subtracted out to obtain the net change
\begin{equation} \label{eq:deltamu}
	\Delta\mu(\sigma)
	= \mu_0 \Bigg(\frac{1-(1+\pi_L\sigma_L+\pi_T\sigma_T)}{1+\pi_L\sigma_L+\pi_T\sigma_T}\Bigg)
	= \mu_0 \frac{-\pi_L\sigma_L-\pi_T\sigma_T}{1+\pi_L\sigma_L+\pi_T\sigma_T}
\end{equation}

\subsection{Oxide Capacitance}
Oxide capacitance changes due to the physical thickness change of the oxide ($t_{ox}$) under strain. This strain ($\Delta t_{ox}/t_{ox0}$) is given by the applied stress divided by the Young's modulus, $E$, of the dielectric, assuming the film is stressed in the elastic regime. Such a relation results in the following expressions:

\begin{equation}
	C_{ox}'(\sigma) = \frac{\epsilon_{ox}}{t_{ox0}+\Delta t_{ox}} = \frac{\epsilon_{ox}}{t_{ox0}(1+\sigma/E_{ox})}
\end{equation}

\begin{equation} \label{eq:deltac}
	\Delta C_{ox}' =  \frac{\epsilon_{ox}}{t_{ox0}}\bigg(\frac{1}{1+\sigma/E_{ox}} - 1\bigg) 
	= C_{ox0}'\bigg(\frac{-\sigma/E_{ox}}{1+\sigma/E_{ox}}\bigg)
\end{equation}

\subsection{Bandgap Modulation}

Strain dependence of the bandgap of silicon ($E_g$) is given by \cite{bardeen_deformation_1950}:
\begin{equation}
	E_g(\xi) = E_{g0} +E_{1g}\xi
\end{equation}
where $E_{1g}$ represents the sensitivity of the bandgap to total strain $\xi$, where 
\begin{equation}
	\xi = \sum_{j = 1}^{3}\sigma_{jj}/E_{jj}. 
\end{equation}

Using this relation, the theoretical dependence of threshold voltage on strain can be derived using a typical expression for extrapolated threshold voltage that is referenced to the approximate onset of strong inversion \cite{tsividis_operation_2010}:

\begin{equation}
	V_T = V_{FB} + \phi_0 + \gamma\sqrt{V_{SB} + \phi_0}
\end{equation}
where $V_{SB}$ is the source to body voltage, $V_{FB}$ is the flat-band voltage governed by the metal-semiconductor work function $\phi_{ms}$ and effective oxide charge per unit area $Q_0'$, by
\begin{equation}
	V_{FB} = \phi_{ms} - Q_0'/C_{ox}',
\end{equation}
$\phi_0$ is a value a few ($2<n<5$ typical) thermal voltages $\phi_t$ above the onset of strong inversion
\begin{equation}
	\phi_0 = 2\phi_f + n\phi_t,
\end{equation}
and
\begin{equation}
	\gamma = \frac{\sqrt{2q\epsilon_sN_A}}{C_{ox}'},
\end{equation}
where  $\epsilon_s$ is the semiconductor permittivity $N_A$ is the semiconductor acceptor doping per unit volume (assuming p-type substrate),
\begin{equation}
	\phi_{ms} = \phi_m-\chi_s-E_g/2 - \phi_f,
\end{equation}
\begin{equation}
	\chi_s = E_{vac}-E_c,
\end{equation}
and 
\begin{equation}
	\phi_f = kT*ln\Big(\frac{N_A}{n_i}\Big),
\end{equation}
with $n_i$ being the intrinsic carrier concentration in the semiconductor ($\approx 10^{10}$ in silicon at room temperature).
Combining all of this together,
\begin{multline} \label{eq:vt_xi}
	V_T(\xi) = \phi_m-\chi_s(\xi)-E_g(\xi)/2 + \phi_f(\xi) - Q_0'(\xi)/C_{ox}'(\xi) \\+ nkT + \frac{\sqrt{2q\epsilon_sN_A(\xi)}}{C_{ox}'(\xi)}\sqrt{V_{SB} + 2\phi_f(\xi) + nkT}
\end{multline}

The dependence of $C_{ox}'$ on strain has already been discussed. The other terms, however, require further examination. Starting with the expression for $n_i$,

\begin{equation}
	n_i = \sqrt{N_cN_v}*exp\Big(\frac{-E_g}{2kT}\Big)
\end{equation}
where $N_{c/v}$ are related to the effective mass of carriers in the semiconductor conduction and valance band as follows:

\begin{equation}
	N_{c/v} = 2*\sqrt[3/2]{\frac{2\pi m_{n/p}^*}{kTh^2}}
\end{equation}
Neglecting the change in $\sqrt{N_cN_v}$ since it will have less impact than the change in the exponential term,

\begin{multline}
	n_i(\xi) \approx \sqrt{N_cN_v}*exp\Big(\frac{-(E_g+\Delta E_g)}{2kT}\Big) 
	= \sqrt{N_cN_v}*exp\Big(\frac{-E_g}{2kT}\Big)exp\Big(\frac{-\Delta E_g}{2kT}\Big) \\
	= n_{i0}*exp\Big(\frac{-\Delta E_g}{2kT}\Big)
\end{multline}

Carrier concentration exhibits a simple dependence on strain as the stretching/compression of the lattice merely changes the number of charges within a fixed volume. To a first order approximation,
\begin{equation}
	N_A(\xi) = \frac{N_{A0}}{1+\xi}
\end{equation}

These expressions, along with bandgap dependence on strain, are now inserted into the expression for $\phi_f$ to obtain
\begin{equation}
	\phi_f(\xi) = kT*ln\Big(\frac{N_{A0}}{n_{i0}}*\frac{exp(\Delta E_g/2kT)}{1+\xi}\Big) 
	= \phi_{f0} + kT*ln\Big(\frac{exp(\Delta E_g/2kT)}{1+\xi}\Big)
\end{equation}
where
\begin{equation}
	\Delta \phi_f = kT*ln\Big(\frac{exp(\Delta E_g/2kT)}{1+\xi}\Big) 
	= \Delta E_g/2 + kT*ln\Big(\frac{1}{1+\xi}\Big)
\end{equation}

Examining the change in threshold voltage as a function of gate dielectric stress and strain in the silicon substrate, the only term that does not have a well defined relation is $Q_0'$. This is because the term is the projection of all parasitic charges in the gate stack onto the semiconductor surface. Thus, while this term is itself an effective sheet charge, it in reality encompasses a mixture of sheet and volume charges which respond differently to strain, the ratio of which is not well defined. This term is thus left alone in the expression below, being defined as an initial ($Q_00'$) and change in ($\Delta Q_0'$) value. Substituting the derived relations into Eq. \ref{eq:vt_xi} and subtracting out the initial threshold voltage value gives the following expression for $\Delta V_T$:

\begin{multline}
	\begin{split}
		\Delta V_T = -\Delta \chi_s + kT*ln\Big(\frac{1}{1+\xi}\Big) 
		+ \Bigg(\frac{\sqrt{2q\epsilon_s\frac{N_{A0}}{1+\xi}}}{C_{ox}'(1+\sigma/E_{ox})} 
		*\sqrt{2\bigg(\phi_{f0}+\Delta E_g/2 + kT*ln\Big(\frac{1}{1+\xi}\Big)\bigg) + V_{SB} + nkT}  \\ 
		- \frac{\sqrt{2q\epsilon_sN_{A0}}}{C_{ox0}'}\sqrt{V_{SB} + 2\phi_{f0} + nkT}\Bigg) 
		- \bigg(\frac{\Delta Q_0'-Q_{00}'\sigma/E_{ox}}{C_{ox}'(1+\sigma/E_{ox})}\bigg)
	\end{split}
\end{multline}

\subsection{Square Law Model}
The source-referenced simplified strong-inversion model with $\alpha=1$ gives the traditional square-law model \cite{tsividis_operation_2010}:
\begin{equation}
	I_d = \mu C_{ox} \frac{W}{L}\Big((V_{GS}-V_T)V_{DS} - \frac{V_{DS}^2}{2}\Big)
\end{equation}
Modified to include the effects of stress on $\mu$, $C_{ox}'$, and $V_T$, this becomes the following:

\begin{multline} %
	I_d(\sigma)	= (\mu_0 C_{ox0}' +\mu_0 \Delta C_{ox}' + C_{ox0}'\Delta \mu + \Delta \mu\Delta C_{ox}')  *\frac{W}{L}\Big((V_{GS}-V_{T0})V_{DS} - \frac{V_{DS}^2}{2}\Big) \\
	-\Delta V_T V_{DS} \frac{W}{L} (\mu_0 C_{ox0}' +\mu_0 \Delta C_{ox}' + C_{ox0}'\Delta \mu + \Delta \mu\Delta C_{ox}')
\end{multline}
where
\begin{multline} %
	\Delta I_d(\sigma)	= (\mu_0 \Delta C_{ox}' + C_{ox0}'\Delta \mu + \Delta \mu\Delta C_{ox}')  *\frac{W}{L}\Big((V_{GS}-V_{T0})V_{DS} - \frac{V_{DS}^2}{2}\Big) \\
	-\Delta V_T V_{DS} \frac{W}{L} (\mu_0 C_{ox0}' +\mu_0 \Delta C_{ox}' + C_{ox0}'\Delta \mu + \Delta \mu\Delta C_{ox}')
\end{multline}

Likewise, in saturation

\begin{equation}\label{eq:idsat_lcm}
	I_{d0} = \mu C_{ox} \frac{W}{2L}\Big((V_{GS}-V_T)^2\big(1+\lambda[V_{DS}-(V_{GS}-V_{T0})]\big)\Big)
\end{equation}
and

\begin{multline}
	\Delta I_d 
	= (\mu_0 \Delta C_{ox}' + C_{ox0}'\Delta \mu + \Delta \mu\Delta C_{ox}')
	*\frac{W}{2L}\Big((V_{GS}-V_T)^2(1+\lambda(V_{DS} - V_{GS} + V_{T}))\Big) \\
	\hspace{13mm}- \frac{W}{2L} (\mu_0 C_{ox0} +\mu_0 \Delta C_{ox} + C_{ox0}\Delta \mu + \Delta \mu\Delta C_{ox}) \Bigg[ \hfill \\ 
	\hspace{27mm}\Delta V_T \Big[\lambda(V_{GS}-V_{T0})^2 - 2(V_{GS}-V_{T0}) *\Big(1+\lambda\big[V_{DS}-(V_{GS}-V_{T0})\big]\Big)\Big] \hfill \\ 
	\hspace{23mm}+ \Delta V_T^2 \Big[\Big(1+\lambda\big[V_{DS}-(V_{GS}-V_{T0})\big]\Big) - 2\lambda(V_{GS}-V_{T0})\Big] \hfill \\
	\hspace{23mm}+ \Delta V_T^3 [\lambda]  \Bigg] \hfill 
\end{multline}

Observing Eq. \ref{eq:deltaid_stressfet} combined with the expressions for change in mobility and capacitance (Eqns. \ref{eq:deltamu} and \ref{eq:deltac}), it can be seen that the initial values for mobility and capacitance can both be factored out of the non-$\Delta V_T$ term, resulting in a change in drain current as follows (for the linear regime): 

\begin{multline}
	\Delta I_d
	= I_{d0}\Bigg(\bigg(\frac{-\sigma/E_{ox}}{1+\sigma/E_{ox}}\bigg) + \bigg(\frac{-\pi_L\sigma_L-\pi_T\sigma_T}{1+\pi_L\sigma_L+\pi_T\sigma_T}\bigg)+ \bigg(\frac{-\sigma/E_{ox}}{1+\sigma/E_{ox}}\bigg)\bigg(\frac{-\pi_L\sigma_L-\pi_T\sigma_T}{1+\pi_L\sigma_L+\pi_T\sigma_T}\bigg)\Bigg) \\
	-\Delta V_T V_{DS} \frac{W}{L} \mu_0 C_{ox0} \Bigg(1+ \bigg(\frac{-\sigma/E_{ox}}{1+\sigma/E_{ox}}\bigg) + \bigg(\frac{-\pi_L\sigma_L-\pi_T\sigma_T}{1+\pi_L\sigma_L+\pi_T\sigma_T}\bigg)+ \bigg(\frac{-\sigma/E_{ox}}{1+\sigma/E_{ox}}\bigg)\bigg(\frac{-\pi_L\sigma_L-\pi_T\sigma_T}{1+\pi_L\sigma_L+\pi_T\sigma_T}\bigg)\Bigg)
\end{multline}
with a similar expression likewise derivable for saturation.

\section{Mobility and Velocity Saturation Effects} \label{sec:mobility}

In sub-micron devices with high lateral fields in the channel region, carrier drift velocity does not exhibit a linear dependence on electric field (constant mobility), but rather saturates as $V_{DS}$ increases \cite{sze_physics_2006}. The first notable deviation in velocity comes from phonon scattering, whereby carriers in motion lose energy to the lattice around them. This can be modeled via an effective temperature for the carriers, with drift velocity decreasing as the temperature differential increases \cite{moll_physics_1964}:

\begin{equation}
	\frac{T_e}{T} = \frac{1}{2}\Bigg[1+ \sqrt{1+ \frac{3\pi}{8}\Big(\frac{\mu_0E}{c_s}\Big)^2}\Bigg]
\end{equation}

\begin{equation}
	v_d = \mu_0E \sqrt{\frac{T}{T_e}}
\end{equation}

Once the field becomes sufficiently large, carriers may obtain enough energy ($E_p$) to emit optical phonons. When this kinetic energy is transferred to optical phonons, the carriers must subsequently accelerate again to regain kinetic energy \cite{mitin_quantum_1999}. The maximum rate at which they may travel is given by setting the kinetic energy equal to optical phonon energy and solving for velocity:

\begin{equation}
	v_s = \sqrt{\frac{2E_p}{m^*}}
\end{equation}

To simplify modeling of velocity saturation in compact models, an empirical formula is typically used to encapsulate these effects into a simple equation:
\begin{equation}
	v_d = \frac{\mu_0 E}{\sqrt[\kappa]{1+(\mu_0 E / v_s)^{\kappa}}}
\end{equation}

For simplicity in deriving an analytical equation for transistor operation with the presence of velocity saturation, a piece-wise approximation may also be utilized \cite{tsividis_operation_2010}:
\begin{equation}
v_d = \begin{cases}
	\frac{\mu |E_x|}{1+|E_x|/E_{sat}} \quad,\quad |E_x| < E_{sat} \\
	v_{sat}\hfill,\quad |E_x| > E_{sat}
\end{cases}
\end{equation}
which, solved at $v_d = v_{sat}$, gives
\begin{equation} \label{eq:esat}
	E_{sat} = \frac{2v_{sat}}{\mu}
\end{equation}

\begin{figure}
	\centering
	\includegraphics[width=0.6\linewidth]{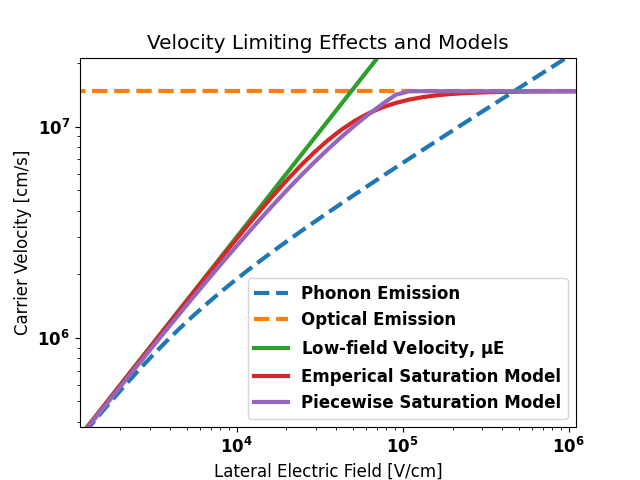}
	\caption{Representative curves for various mobility models. $\mu$=300 cm$^2$/V*s, $c_s \approx 8400$ m/s, $E_p$ = 63 meV, $m^*$=0.98, $\kappa$=2.}
	\label{fig:vsatrelations}
\end{figure}

A comparison of these drift velocity models is shown in Fig. \ref{fig:vsatrelations}, with the understanding that the absolute values of these effects are influenced by doping, device orientation, strain engineering, temperature, and more. For silicon, the two most important phenomena limiting low-field mobility are acoustic-phonon interactions and ionized impurities, with overall mobility calculated as the harmonic mean (Matthiessen rule) of all individual contributions \cite{sze_physics_2006}.

\begin{equation} \label{eq:muph}
	\mu_{ph} = \frac{\sqrt{8\pi}q\hbar^4C_l}{3E^2_{ds}m_c^{*5/2}(kT)^{3/2}}
\end{equation}

\begin{equation} \label{eq:mui}
	\mu_i = \frac{64\sqrt{\pi}\epsilon_s^2(2kT)^{3/2}}{N_Iq^3m^{*1/2}}\Bigg[ln\Big[1+\Big(\frac{12\pi\epsilon_skT}{q^2 N_I^{1/3}}\Big)^2\Big]\Bigg]^{-1}
\end{equation}

\begin{figure}
	\centering
	\includegraphics[width=0.6\linewidth]{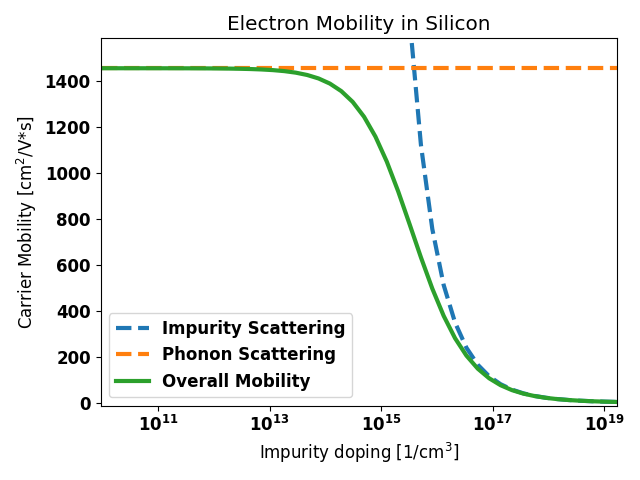}
	\caption{Combination of phonon-limited mobility and impurity-limited mobility via Mattiessen rule results in a familiar shape for electrons in Si. Exact relations are presented in Equations \ref{eq:muph} and \ref{eq:mui}. $T$=300 K, $m^*$=0.98, $C_l$=165 GPa, $\rho$=2,329 kg/m$^3$, $\epsilon_s$=11.9$\epsilon_0$, $E_{ds}$=-2.7 eV}
	\label{fig:mobilityrelations}
\end{figure}

Of note is the fact that both mobility and optical phonon emission depend on effective mass, meaning the change in transistor current due to an effect that alters effective mass (such as acoustic deformation of the crystal lattice) will not be subject to velocity saturation like its DC component. Thus, the approach taken in this work is to use a compound model, with unsaturated long-channel expressions used to derive the relative AC change in transistor current and a saturation model, developed hereafter, to obtain the steady-state drain current used for the unstrained initial condition.

\section{Velocity Saturated Square Law Model} \label{sec:scm}
Utilizing the previously mentioned piece-wise model for carrier drift velocity leads to a modified version of the long channel model:
\begin{equation}
	I_d = \begin{cases}
			\mu C_{ox} \cfrac{W}{L}\cfrac{1}{1+\cfrac{V_{DS}}{E_{sat}L}}\Big((V_{GS}-V_T)V_{DS} - \frac{V_{DS}^2}{2}\Big) \quad,\quad V_{DS} < V_{DSsat} \\ \\
		 C_{ox} W v_{sat}\cfrac{(V_{GS}-V_T)^2}{(V_{GS}-V_T)+E_{sat}(L-\Delta L)} \hfill,\quad V_{DS} > V_{DSsat}
	\end{cases}
\end{equation}
where
\begin{equation}
	V_{DSsat} = \frac{(V_{GS}-V_T)E_{sat}L}{(V_{GS}-V_T)+E_{sat}L}.
\end{equation}
In the long channel limit, $V_{DSsat}$ tends towards $V_{GS}-V_T$ and Equation \ref{eq:esat} can be used to substitute in for $v_{sat}$. Furthermore, linearizing the channel length modulation via Taylor expansion
\begin{equation}
	\frac{1}{1-\Delta L/L} \approx 1+\Delta L/L
\end{equation}
and taking
\begin{equation}
	\Delta L/L = \lambda(V_{DS}-V_{DSsat})
\end{equation}
gives the original long channel expression, shown in Equation \ref{eq:idsat_lcm}.

The end result of this is that the DC saturation current no longer scales quadraticly with $V_{GS}-V_T$ but rather linearly reduces the overall current. While the simultaneous change in mobility and saturation velocity  with stress prevents AC current waveforms from clipping, this does reduces the magnitude of the mobility and oxide capacitance modulation terms that are linearly proportional to steady-state drain current. 

\pagebreak

\section{Model Fit Report} \label{sec:fit_report}
\begin{lstlisting}
[[Model]]
    Model(Id_DC_lmfit)
[[Fit Statistics]]
    # fitting method   = basinhopping
    # function evals   = 1309
    # data points      = 81
    # variables        = 6
    chi-square         = 6.2464e-08
    reduced chi-square = 8.3285e-10
    Akaike info crit   = -1687.63316
    Bayesian info crit = -1673.26646
##  Warning: uncertainties could not be estimated:
    eot:    at initial value
    eot:    at boundary
    Weff0:  at boundary
[[Variables]]
    u0:     0.00788908 +/-        nan (nan%) (init = 0.03)
    eot:    1.6054e-09 +/-        nan (nan%) (init = 1e-09)
    Vt0:    0.20593644 +/- 0.00395160 (1.92%) (init = 0.3)
    vsat:   183092.801 +/- 19456.8380 (10.63%) (init = 200000)
    clm:    0.47366497 +/-        nan (nan%) (init = 0.3)
    Weff0:  3.0000e-06 +/-        nan (nan%) (init = 3.2e-06)


[[Model]]
    Model(delId_vsat_lmfit)
[[Fit Statistics]]
    # fitting method   = leastsq
    # function evals   = 67
    # data points      = 81
    # variables        = 3
    chi-square         = 2.4219e-12
    reduced chi-square = 3.1050e-14
    Akaike info crit   = -2516.41319
    Bayesian info crit = -2509.22984
[[Variables]]
    stressx:  1004021.25 +/- 1186898.72 (118.21%) (init = 1000000)
    stressy:  10000.0691 +/- 1072828.08 (10728.21%) (init = 1000000)
    stressz:  10000.0001 +/- 76609.1709 (766.09%) (init = 1000000)
[[Correlations]] (unreported correlations are < 0.100)
    C(stressx, stressy) = -0.998
    C(stressx, stressz) = -0.294
    C(stressy, stressz) =  0.241
\end{lstlisting}

\pagebreak
\section{Supplemental Figures} \label{sec:sup_figs}

\begin{figure*}[h!]
	\centering
	\includegraphics[width=0.9\linewidth]{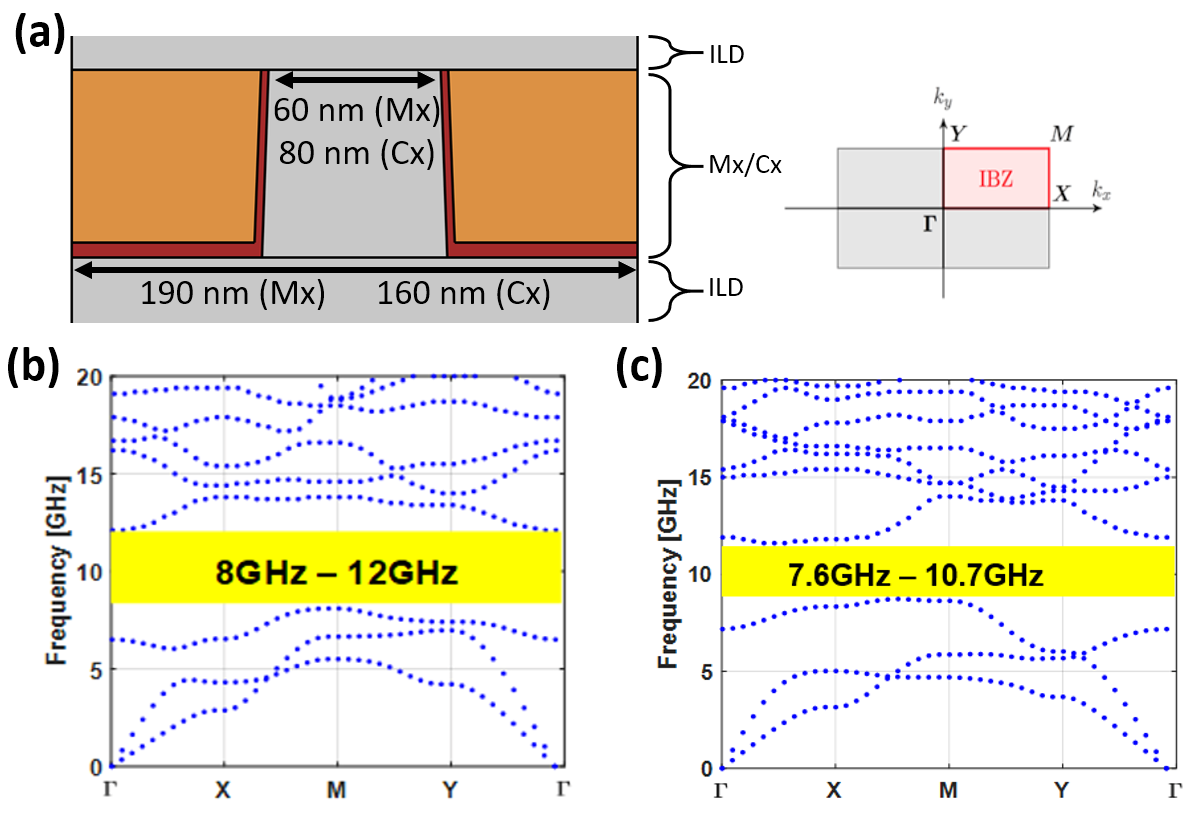}
	\caption{\textbf{(a)} Schematic of the unit cell for the periodic PnC structures, including the metal layer as well as inter-layer dielectric (ILD) films present between each metal layer. The orange region represents the metal whereas the surrounding layers are dielectric. \textbf{(b)} corresponding dispersion relationship with metal dimension of 130 nm with 60 nm gap (Mx). A complete bandgap is obtained between 8 GHz and 12 GHz. \textbf{(c)} highlights the dispersion relationship for PnC structures designed in Cx metal layers (80 nm metal, 80 nm gap). The corresponding bandgap is obtained between 7.6 GHz and 10.7 GHz.}
	\label{fig:DR}
\end{figure*}

\begin{figure*}[h!]
	\centering
	\includegraphics[width=0.7\linewidth]{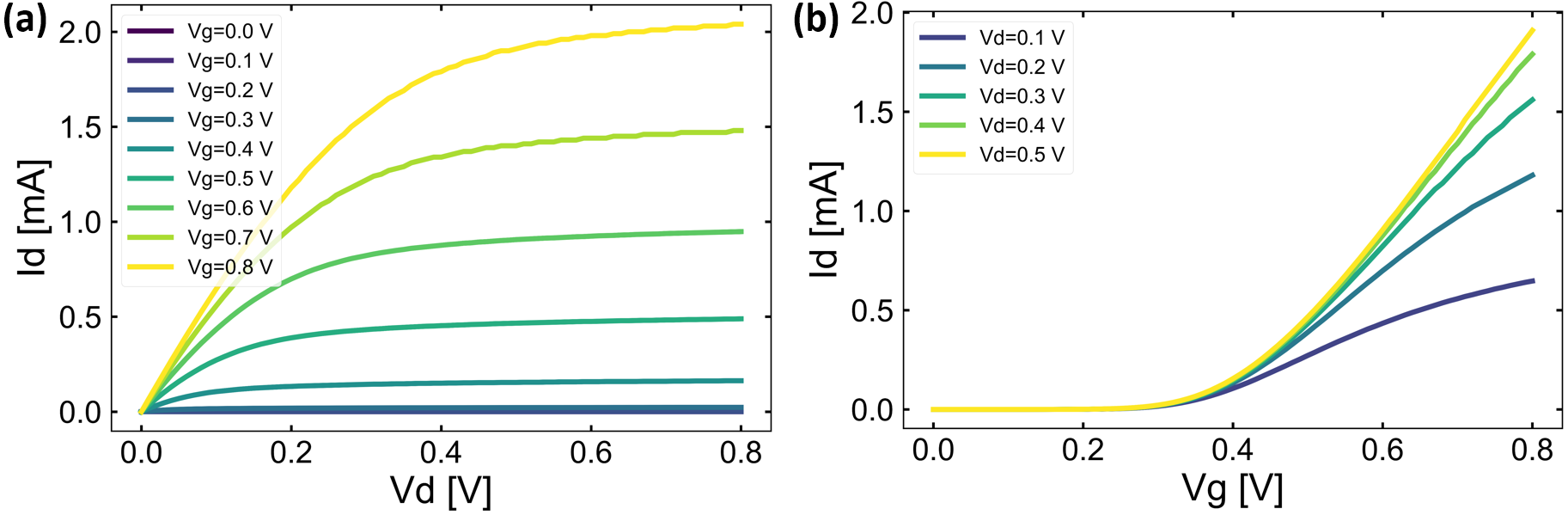}
	\caption{Experimental \textbf{(a)} family of curves and \textbf{(b)} $\mathrm{I_D-V_G}$ DC bias currents for A1 resonator with 80 nm gate length. Drain currents reported are high due to the large width of the sense transistors, which comprise of 40 fins in parallel.}
	\label{fig:dcresults}
\end{figure*}

\begin{figure*}[h!]
	\centering
	\includegraphics[width=0.85\linewidth]{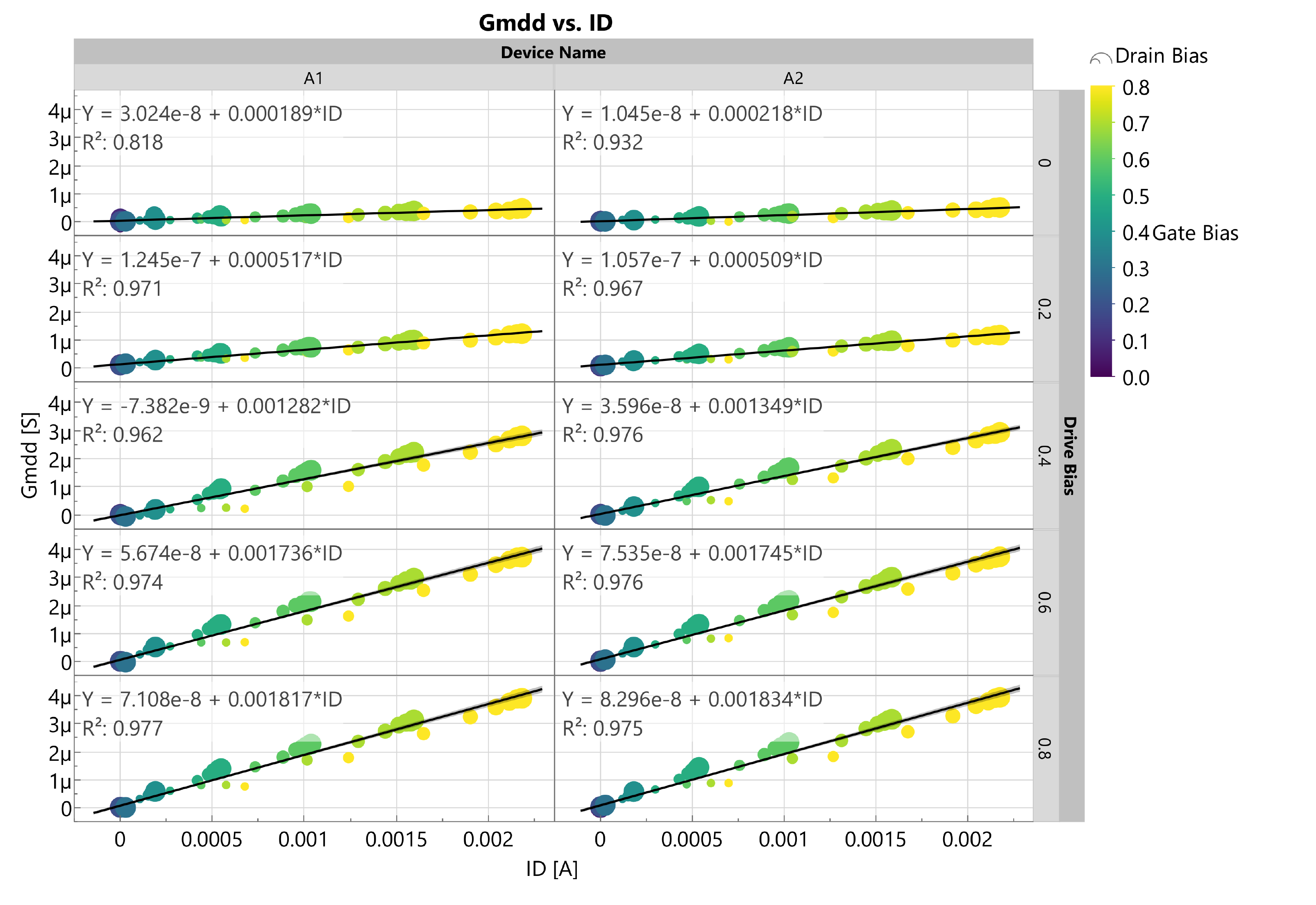}
	\caption{Trend in differential transconductance of largest mode ($\approx$11.75 GHz) for A1 and A2 resonators (80 nm gate length, Mx and Cx phononic crystal confinement). $g_{mdd}$ (Y-axis) is plotted against DC bias current of the sense transistor (X-axis), tiled by device (X tiles) and drive bias (Y tiles). Gate bias is shown by marker color (blue to yellow), with gate biases of 0.3 V and below overlapping near zero drain current (subthreshold). Drain bias is represented with marker size, moving from small to large as bias increases. The large cluster of points with equal gate biases and similar drain currents indicate saturation mode operation for those bias conditions with channel length modulation. Peak $g_{mdd}$ increases as a function of drive bias and is largely dependent on drain current of sense transistor, although efficiency drops as gate voltage increases. Very low drain biases also tend to be less efficient than saturation mode biasing for a given device current.}
	\label{fig:gmdd_fit}
\end{figure*}

\begin{figure}[h!]
	\centering
	\includegraphics[width=0.5\linewidth]{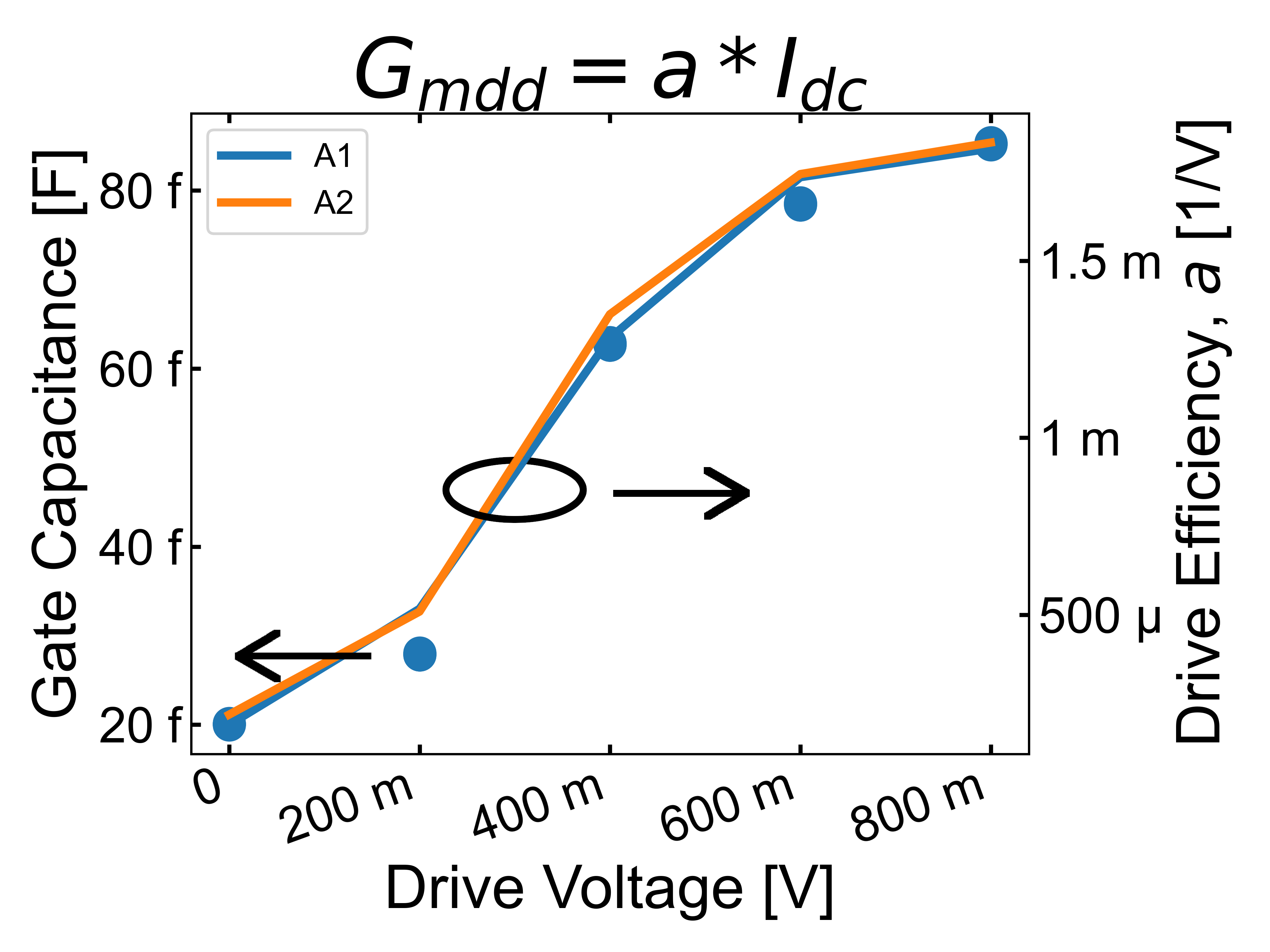}
	\caption{Observed drive efficiency of resonators from the linear fits shown in Fig. \ref{fig:gmdd_fit} plotted alongside extracted capacitance from drive transducers vs drive bias on the X-axis indicating that transduction efficiency is dominated by the large change in capacitance in these devices over the small allowable voltage range.}
	\label{fig:driveresults}
\end{figure}

\begin{figure}
	\centering
	\includegraphics[width=0.8\linewidth]{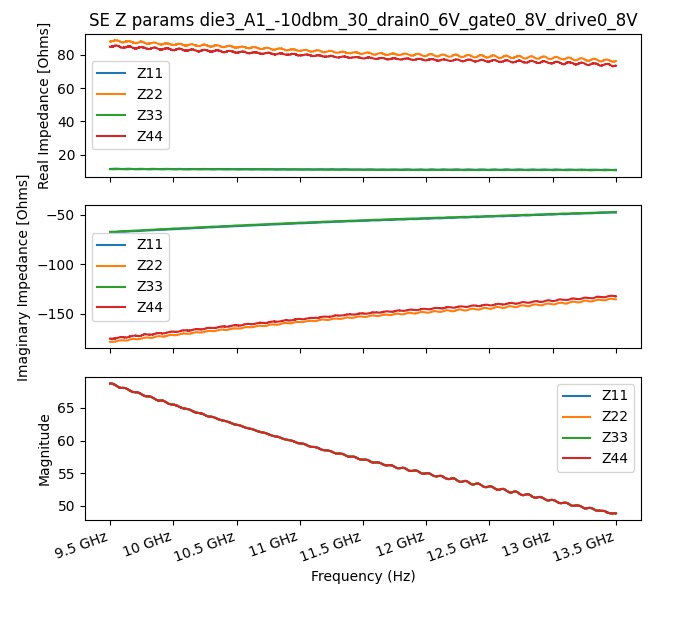}
	\caption{Experimentally measured resonator impedance at input (ports 1 and 3) and output (ports 2 and 4), including input ESD diodes and device to pad routing. }
	\label{fig:impedance}
\end{figure}

\FloatBarrier

\section*{Acknowledgment}

The authors would like to thank GlobalFoundries for funding the tapeout of the resonators, as well as the DARPA UPSIDE (contract \# 12405-301701-DS) and MIDAS (contract \# FA8650-18-1-7904) programs for funding the device measurement and analysis.

\section*{Author Contribution}
J. Anderson performed device measurements, analysis, electrical modeling, and led manuscript preparation. Y. He wrote the sections on phononic confinement and cavity termination schemes and performed simulations related to device acoustic design. B. Bahr designed the devices and managed the tapeout of the resonators measured, as well as construction of an initial COMSOL model for device simulation. D. Weinstein provided feedback and guidance to the authors across device design, measurement, and manuscript preparation.

\bibliographystyle{IEEEtran}
\bibliography{biblio.bib}

\begin{thebibliography}{10}
\providecommand{\url}[1]{#1}
\csname url@samestyle\endcsname
\providecommand{\newblock}{\relax}
\providecommand{\bibinfo}[2]{#2}
\providecommand{\BIBentrySTDinterwordspacing}{\spaceskip=0pt\relax}
\providecommand{\BIBentryALTinterwordstretchfactor}{4}
\providecommand{\BIBentryALTinterwordspacing}{\spaceskip=\fontdimen2\font plus
\BIBentryALTinterwordstretchfactor\fontdimen3\font minus
  \fontdimen4\font\relax}
\providecommand{\BIBforeignlanguage}[2]{{%
\expandafter\ifx\csname l@#1\endcsname\relax
\typeout{** WARNING: IEEEtran.bst: No hyphenation pattern has been}%
\typeout{** loaded for the language `#1'. Using the pattern for}%
\typeout{** the default language instead.}%
\else
\language=\csname l@#1\endcsname
\fi
#2}}
\providecommand{\BIBdecl}{\relax}
\BIBdecl

\bibitem{wang_film_2020}
J.~Wang, M.~Park, S.~Mertin, T.~Pensala, F.~Ayazi, and A.~Ansari, ``A {Film}
  {Bulk} {Acoustic} {Resonator} {Based} on {Ferroelectric} {Aluminum}
  {Scandium} {Nitride} {Films},'' \emph{Journal of Microelectromechanical
  Systems}, vol.~29, no.~5, pp. 741--747, Oct. 2020, conference Name: Journal
  of Microelectromechanical Systems.

\bibitem{he2020tunable}
Y.~He, B.~Bahr, M.~Si, P.~Ye, and D.~Weinstein, ``A tunable ferroelectric based
  unreleased rf resonator,'' \emph{Microsystems \& Nanoengineering}, vol.~6,
  no.~1, pp. 1--7, 2020.

\bibitem{lu_enabling_2020}
R.~Lu, Y.~Yang, S.~Link, and S.~Gong, ``Enabling {Higher} {Order} {Lamb} {Wave}
  {Acoustic} {Devices} {With} {Complementarily} {Oriented} {Piezoelectric}
  {Thin} {Films},'' \emph{Journal of Microelectromechanical Systems}, vol.~29,
  no.~5, pp. 1332--1346, Oct. 2020, conference Name: Journal of
  Microelectromechanical Systems.

\bibitem{assylbekova_11_2020}
M.~Assylbekova, G.~Chen, G.~Michetti, M.~Pirro, L.~Colombo, and M.~Rinaldi,
  ``11 {GHz} {Lateral}-{Field}-{Excited} {Aluminum} {Nitride}
  {Cross}-{Sectional} {Lamé} {Mode} {Resonator},'' in \emph{2020 {Joint}
  {Conference} of the {IEEE} {International} {Frequency} {Control} {Symposium}
  and {International} {Symposium} on {Applications} of {Ferroelectrics}
  ({IFCS}-{ISAF})}, Jul. 2020, pp. 1--4, iSSN: 2375-0448.

\bibitem{yang_fifty_2015}
S.~Yang and L.~Hanzo, ``Fifty {Years} of {MIMO} {Detection}: {The} {Road} to
  {Large}-{Scale} {MIMOs},'' \emph{IEEE Communications Surveys Tutorials},
  vol.~17, no.~4, pp. 1941--1988, 2015, conference Name: IEEE Communications
  Surveys Tutorials.

\bibitem{fedder_technologies_2008}
G.~K. Fedder, R.~T. Howe, T.~K. Liu, and E.~P. Quevy, ``Technologies for
  {Cofabricating} {MEMS} and {Electronics},'' \emph{Proceedings of the IEEE},
  vol.~96, no.~2, pp. 306--322, Feb. 2008, conference Name: Proceedings of the
  IEEE.

\bibitem{chen_cmos-integrated_2019}
C.~Chen, M.~Li, A.~A. Zope, and S.~Li, ``A {CMOS}-{Integrated} {MEMS}
  {Platform} for {Frequency} {Stable} {Resonators}-{Part} {I}: {Fabrication},
  {Implementation}, and {Characterization},'' \emph{Journal of
  Microelectromechanical Systems}, vol.~28, no.~5, pp. 744--754, Oct. 2019,
  conference Name: Journal of Microelectromechanical Systems.

\bibitem{riverola_single-resonator_2017}
M.~Riverola, G.~Sobreviela, F.~Torres, A.~Uranga, and N.~Barniol,
  ``Single-{Resonator} {Dual}-{Frequency} {BEOL}-{Embedded} {CMOS}-{MEMS}
  {Oscillator} {With} {Low}-{Power} and {Ultra}-{Compact} {TIA} {Core},''
  \emph{IEEE Electron Device Letters}, vol.~38, no.~2, pp. 273--276, Feb. 2017,
  conference Name: IEEE Electron Device Letters.

\bibitem{bahr_phononic_2014}
B.~Bahr, R.~Marathe, and D.~Weinstein, ``Phononic crystals for acoustic
  confinement in {CMOS}-{MEMS} resonators,'' in \emph{2014 {IEEE}
  {International} {Frequency} {Control} {Symposium} ({FCS})}, May 2014, pp.
  1--4, iSSN: 2327-1949.

\bibitem{bahr_theory_2015}
------, ``Theory and {Design} of {Phononic} {Crystals} for {Unreleased}
  {CMOS}-{MEMS} {Resonant} {Body} {Transistors},'' \emph{Journal of
  Microelectromechanical Systems}, vol.~24, no.~5, pp. 1520--1533, Oct. 2015,
  conference Name: Journal of Microelectromechanical Systems.

\bibitem{bahr_32ghz_2018}
B.~Bahr, Y.~He, Z.~Krivokapic, S.~Banna, and D.~Weinstein, ``{32GHz}
  resonant-fin transistors in 14nm {FinFET} technology,'' in \emph{2018 {IEEE}
  {International} {Solid} - {State} {Circuits} {Conference} - ({ISSCC})}, Feb.
  2018, pp. 348--350, iSSN: 2376-8606.

\bibitem{razavi_jitter-power_2021}
\BIBentryALTinterwordspacing
B.~Razavi, ``\BIBforeignlanguage{en}{Jitter-{Power} {Trade}-{Offs} in
  {PLLs}},'' \emph{\BIBforeignlanguage{en}{IEEE Transactions on Circuits and
  Systems I: Regular Papers}}, vol.~68, no.~4, pp. 1381--1387, Apr. 2021.
  [Online]. Available: \url{https://ieeexplore.ieee.org/document/9354431/}
\BIBentrySTDinterwordspacing

\bibitem{nikonov_convolutional_nodate}
D.~E. Nikonov, I.~A. Young, and G.~I. Bourianoff,
  ``\BIBforeignlanguage{en}{Convolutional {Networks} for {Image} {Processing}
  by {Coupled} {Oscillator} {Arrays}},'' p.~23.

\bibitem{noauthor_comsol_nodate}
\BIBentryALTinterwordspacing
``{COMSOL}: {Multiphysics} {Software} for {Optimizing} {Designs}.'' [Online].
  Available: \url{https://www.comsol.com/}
\BIBentrySTDinterwordspacing

\bibitem{globalfoundries_12lp_2018}
\BIBentryALTinterwordspacing
{GlobalFoundries}, ``{12LP}: 12nm {FinFET} {Technology},'' 2018. [Online].
  Available:
  \url{https://www.globalfoundries.com/sites/default/files/product-briefs/pb-12lp-11-web.pdf}
\BIBentrySTDinterwordspacing

\bibitem{trentzsch_28nm_2016}
M.~Trentzsch, S.~Flachowsky, R.~Richter, J.~Paul, B.~Reimer, D.~Utess,
  S.~Jansen, H.~Mulaosmanovic, S.~Müller, S.~Slesazeck, J.~Ocker, M.~Noack,
  J.~Müller, P.~Polakowski, J.~Schreiter, S.~Beyer, T.~Mikolajick, and
  B.~Rice, ``A 28nm {HKMG} super low power embedded {NVM} technology based on
  ferroelectric {FETs},'' in \emph{2016 {IEEE} {International} {Electron}
  {Devices} {Meeting} ({IEDM})}.\hskip 1em plus 0.5em minus 0.4em\relax IEEE,
  Dec. 2016, pp. 11.5.1--11.5.4.

\bibitem{he2019switchable}
Y.~He, B.~Bahr, M.~Si, P.~Ye, and D.~Weinstein, ``Switchable mechanical
  resonance induced by hysteretic piezoelectricity in ferroelectric
  capacitors,'' in \emph{2019 20th International Conference on Solid-State
  Sensors, Actuators and Microsystems \& Eurosensors XXXIII (TRANSDUCERS \&
  EUROSENSORS XXXIII)}.\hskip 1em plus 0.5em minus 0.4em\relax IEEE, 2019, pp.
  717--720.

\bibitem{rawat_analysis_2021}
U.~Rawat, B.~Bahr, and D.~Weinstein, ``Analysis and {Modeling} of a 11.8 {GHZ}
  {Fin} {Resonant} {Body} {Transistor} in a 14 nm {FinFET} {CMOS} {Process},''
  \emph{Under Review}, 2021.

\bibitem{meade1995photonic}
R.~Meade, J.~N. Winn, and J.~Joannopoulos, ``Photonic crystals: Molding the
  flow of light,'' \emph{Pinceton Univ. Press}, 1995.

\bibitem{bahr2015theory}
B.~Bahr, R.~Marathe, and D.~Weinstein, ``Theory and design of phononic crystals
  for unreleased cmos-mems resonant body transistors,'' \emph{Journal of
  Microelectromechanical systems}, vol.~24, no.~5, pp. 1520--1533, 2015.

\bibitem{noauthor_bsim_nodate}
\BIBentryALTinterwordspacing
``\BIBforeignlanguage{en-US}{{BSIM} {Group}}.'' [Online]. Available:
  \url{http://bsim.berkeley.edu/}
\BIBentrySTDinterwordspacing

\bibitem{tsividis_operation_2010}
Y.~Tsividis and C.~McAndrew, \emph{\BIBforeignlanguage{English}{Operation and
  {Modeling} of the {MOS} {Transistor}}}, 3rd~ed.\hskip 1em plus 0.5em minus
  0.4em\relax New York: Oxford University Press, Sep. 2010.

\bibitem{taur_continuous_2004}
Y.~Taur, X.~Liang, W.~Wang, and H.~Lu, ``A continuous, analytic drain-current
  model for {DG} {MOSFETs},'' \emph{IEEE Electron Device Letters}, vol.~25,
  no.~2, pp. 107--109, Feb. 2004, conference Name: IEEE Electron Device
  Letters.

\bibitem{bahr_optimization_2016}
\BIBentryALTinterwordspacing
B.~Bahr, L.~Daniel, and D.~Weinstein, ``\BIBforeignlanguage{en}{Optimization of
  unreleased {CMOS}-{MEMS} {RBTs}},'' in \emph{\BIBforeignlanguage{en}{2016
  {IEEE} {International} {Frequency} {Control} {Symposium} ({IFCS})}}.\hskip
  1em plus 0.5em minus 0.4em\relax New Orleans, LA, USA: IEEE, May 2016, pp.
  1--4. [Online]. Available: \url{http://ieeexplore.ieee.org/document/7563592/}
\BIBentrySTDinterwordspacing

\bibitem{marathe_resonant_2014}
R.~Marathe, B.~Bahr, W.~Wang, Z.~Mahmood, L.~Daniel, and D.~Weinstein,
  ``Resonant {Body} {Transistors} in {IBM}'s 32 nm {SOI} {CMOS} {Technology},''
  \emph{Journal of Microelectromechanical Systems}, vol.~23, no.~3, pp.
  636--650, Jun. 2014, conference Name: Journal of Microelectromechanical
  Systems.

\bibitem{chen_high-q_2019}
G.~Chen and M.~Rinaldi, ``High-{Q} {X} {Band} {Aluminum} {Nitride} {Combined}
  {Overtone} {Resonators},'' in \emph{2019 {Joint} {Conference} of the {IEEE}
  {International} {Frequency} {Control} {Symposium} and {European} {Frequency}
  and {Time} {Forum} ({EFTF}/{IFC})}, Apr. 2019, pp. 1--3, iSSN: 2327-1949.

\bibitem{yang_1060-ghz_2020}
Y.~Yang, R.~Lu, L.~Gao, and S.~Gong, ``10–60-{GHz} {Electromechanical}
  {Resonators} {Using} {Thin}-{Film} {Lithium} {Niobate},'' \emph{IEEE
  Transactions on Microwave Theory and Techniques}, vol.~68, no.~12, pp.
  5211--5220, Dec. 2020, conference Name: IEEE Transactions on Microwave Theory
  and Techniques.

\bibitem{abouyoussef_quad_2018}
M.~S. Abouyoussef, A.~M. El-Tager, and H.~El-Ghitani, ``Quad spiral microstrip
  resonator with high quality factor,'' in \emph{2018 35th {National} {Radio}
  {Science} {Conference} ({NRSC})}, Mar. 2018, pp. 6--13.

\bibitem{anderson_pymeasrf_2019}
\BIBentryALTinterwordspacing
J.~Anderson and D.~Weinstein, ``{PyMeasRF}: {Automating} {RF} {Device}
  {Measurements} {Using} {Python},'' Jul. 2019. [Online]. Available:
  \url{https://zenodo.org/record/3346201}
\BIBentrySTDinterwordspacing

\bibitem{anderson_pymeasrf_2020}
\BIBentryALTinterwordspacing
J.~Anderson, ``Pymeasrf,'' Jan. 2020, original-date: 2018-01-04T01:57:30Z.
  [Online]. Available: \url{https://github.com/JAnderson419/PyMeasRF}
\BIBentrySTDinterwordspacing

\bibitem{scikit-rf_developers_scikit-rf_2020}
\BIBentryALTinterwordspacing
{Scikit-rf Developers}, ``Scikit-rf,'' 2020. [Online]. Available:
  \url{http://scikit-rf.org/index.html}
\BIBentrySTDinterwordspacing

\bibitem{noauthor_lmfit_2020}
\BIBentryALTinterwordspacing
``{LMFIT}: {Non}-{Linear} {Least}-{Squares} {Minimization} and
  {Curve}-{Fitting} for {Python},'' 2020. [Online]. Available:
  \url{https://lmfit.github.io/lmfit-py/}
\BIBentrySTDinterwordspacing

\bibitem{noauthor_dask_2020}
\BIBentryALTinterwordspacing
``Dask: {Scalable} analytics in {Python},'' 2020. [Online]. Available:
  \url{https://dask.org/}
\BIBentrySTDinterwordspacing

\bibitem{noauthor_jmp_2019}
\BIBentryALTinterwordspacing
``jmp,'' 2019. [Online]. Available: \url{https://www.jmp.com/en\_us/home.html}
\BIBentrySTDinterwordspacing

\bibitem{bardeen_deformation_1950}
\BIBentryALTinterwordspacing
J.~Bardeen and W.~Shockley, ``Deformation {Potentials} and {Mobilities} in
  {Non}-{Polar} {Crystals},'' \emph{Physical Review}, vol.~80, no.~1, pp.
  72--80, Oct. 1950. [Online]. Available:
  \url{https://link.aps.org/doi/10.1103/PhysRev.80.72}
\BIBentrySTDinterwordspacing

\bibitem{sze_physics_2006}
S.~M. Sze and K.~K. Ng, \emph{\BIBforeignlanguage{English}{Physics of
  {Semiconductor} {Devices}}}, 3rd~ed.\hskip 1em plus 0.5em minus 0.4em\relax
  Hoboken, N.J: Wiley-Interscience, Oct. 2006.

\bibitem{moll_physics_1964}
J.~L. Moll, \emph{\BIBforeignlanguage{English}{Physics of
  {Semiconductors}}}.\hskip 1em plus 0.5em minus 0.4em\relax McGraw-Hill, Jan.
  1964.

\bibitem{mitin_quantum_1999}
V.~Mitin, V.~V. Mitin, V.~Kochelap, and M.~A. Stroscio,
  \emph{\BIBforeignlanguage{en}{Quantum {Heterostructures}: {Microelectronics}
  and {Optoelectronics}}}.\hskip 1em plus 0.5em minus 0.4em\relax Cambridge
  University Press, Jul. 1999, google-Books-ID: Wzo4IdxS48oC.

\end{thebibliography}

\end{document}